\documentclass[12pt]{iopart}
\usepackage{iopams,bm,url,amssymb,graphicx,color}

\begin{document}

\title[TDDFT of silver clusters and shells]{
Optical response of silver clusters and their hollow shells
from Linear-Response TDDFT}

\author{Peter Koval and Federico Marchesin}
\address{Centro de F\'isica de Materiales, 
Centro Mixto CSIC-UPV/EHU and Donostia International Physics Center (DIPC), 
Paseo Manuel de Lardizabal 5, 20018 Donostia-San Sebasti\'an, Spain}
\ead{koval.peter@gmail.com}

\author{Dietrich Foerster}
\address{Laboratoire Ondes et Mati\`e{}re d'Aquitaine, 
Universit\'e de Bordeaux 1, 351 Cours de la Liberation,
   33405 Talence, France;}

\author{Daniel S\'anchez-Portal}
\address{Centro de F\'isica de Materiales, 
Centro Mixto CSIC-UPV/EHU and Donostia International Physics Center (DIPC), 
Paseo Manuel de Lardizabal 5, 20018 Donostia-San Sebasti\'an, Spain}
\ead{sqbsapod@ehu.es}

\begin{abstract}
We present a study of the optical response of compact and hollow 
icosahedral clusters containing up to 868 silver atoms by means of
time-dependent density functional theory. We have studied the dependence
on size and morphology of both the sharp plasmonic resonance at 3--4 eV
(originated mainly from $sp$-electrons), and the less studied broader
feature appearing in the 6--7 eV range (interband transitions).
An analysis of the effect of structural relaxations, as well as the
choice of exchange correlation functional (local density versus 
generalized gradient approximations)
both in the ground state and optical response calculations is also
presented. We have further analysed the role of the different atom
layers (surface versus inner layers) and the different orbital symmetries
on the absorption cross-section for energies up to 8 eV. We have also
studied the dependence on the number of atom layers in hollow structures.
Shells formed by a single layer of atoms show a pronounced red shift of
the main plasmon resonances that, however, rapidly converge to those of
the compact structures as the number of layers is increased. The methods
used to obtain these results are also carefully discussed. 
Our methodology is based on the use of localized basis (atomic orbitals, 
and atom-centered- and dominant- product functions), which bring several 
computational advantages related to their relatively small size and 
the sparsity of the resulting matrices. Furthermore, the use of basis 
sets of atomic orbitals also brings the possibility to extend some of 
the standard population analysis tools (e.g., Mulliken population 
analysis) to the realm of optical excitations. Some examples of these 
analyses are described in the present work.
\end{abstract}

%
\vspace{2pc}

%
\submitto{\JPCM}
%
%
%

%
%
%
\section{Introduction}

The scientific interest in metallic clusters persists after many decades
of extensive study \cite{Heer:1993}. Among the other metals, silver
is one of the most popular materials for production of clusters.
The primer interest in silver clusters originates from 
the reduced chemical activity of silver \cite{Silver-encyclopedia:2015},
the best electron conducting properties, a sharp plasmonic resonance 
at a relatively low frequency and an affordable price of the metal.
Silver clusters are widely used in various applications
\cite{Brack-rev-mod-phys:1993,Supported-AgNP:2014,Recent-plasmonics-book:2015,
appl-np-pv:2016}
and in the past had also been widely theoretically studied  
\cite{Mie-book-Kreibig:1995,dda-Schatz:2004,Pushka:2010,Rabilloud:2015,
gpaw-prop:2015}.
It is experimentally established that silver nano-particles (NP) 
of diameters 10 to 80 nm exhibit a surface-plasmon resonance in a wide
frequency range from blue to green visible light depending on
their size. Moreover, by controlling the shape of silver NP 
one can further lower the resonance frequency 
\cite{shape-exp:2005,theo-mie-shells:2006,sigmaaldrich:2015}.
Recently, also subnanometric noble metal clusters have been produced
and received large attention in the research community \cite{review-qc:2014}.
The use of subnanometric silver clusters has been demonstrated
in catalysis, \cite{appl-catalysis:2006,appl-catalysis:2011}
in chemical sensing, \cite{appl-detect:2011,appl-detect:2012,appl-detect-halide:2012}
and in surface-enchanced Raman scattering 
\cite{appl-sers:2013,review-sers-theo:2015}, among the other fields.

From a theoretical point of view, 
the plasmonic properties of large silver particles (20 nm and larger)
can be satisfactory described by classical Mie theory 
\cite{Mie-book-Kreibig:1995}. Further extensions of the classical description
include the discrete-dipole approximation \cite{dda-Schatz:2004} and the
finite-difference time-domain techniques \cite{fdtd-Norlander:2007}
that are capable of describing the response of classical objects of any shape.
For smaller clusters, of an effective diameter less than 3 nm 
($\sim$ 600 atoms), the classical theory cannot give a rigorous description 
\cite{dda-Schatz:2004,Charle:1998,tb-atomistic:2013,tddft_iter_plasmons_na380} 
because the atomistic details can significantly alter the classically
averaged picture and it is necessary to make an accurate description
of the electronic distribution and scattering at surfaces.
Therefore, quantum mechanical methods had been widely 
applied in studies of silver clusters 
\cite{Pushka:2010,Rabilloud:2015,gpaw-prop:2015,tddft-clusters-schatz:2008}, 
silver shells \cite{Weissker:2014, Stener-shells:2014}, and 
alloys \cite{theo-alloys:2011,Stener-Ag-Pt:2014}.
In this paper, we will focus on relatively large icosahedral clusters
containing up to 561 atoms and shells of up to 868 atoms 
that were not addressed in the past.

From a methodological perspective, there are early studies 
of silver clusters of different sizes and shapes using
H\"u{}ckel models \cite{Hueckel:1973}, tight-binding techniques 
\cite{tb:2001,tb-atomistic:2013}, density-functional theory (DFT) 
\cite{dft:2002,tddft-clusters-schatz:2008,dft-rods:2009,
octopus-ag-au:2013,small-ag-cu:2014},
time-dependent DFT (TDDFT) 
\cite{gpaw-prop:2015,Ag13-lda-analysis:2008,
xc-contr-ag13-ag55:2010,octopus-ag-au:2013}
and many-body perturbation theory \cite{mbpt:2003,abinitio-cis:2012}.
In this work, we apply linear-response TDDFT using a
linear combination of atomic orbitals (LCAO) as a basis set 
to describe the electronic states of the clusters.
Our method \cite{iter_method,pssb:2010} has
been recently enhanced in several respects. 
Besides the generality of the geometries and chemical species
characteristic of \textit{ab-initio} methods (in the present 
case our linear-response solver is coupled to the 
SIESTA method \cite{siesta:2002, siesta:2008}),
the main advantage of the method is its computational efficiency
that stems from the use of an iterative scheme to compute the optical
response, as well as an efficient basis to express the products
of atomic orbitals. Importantly, the number of iterations does not 
scale significantly with the size of the system studied.
The frequency-by-frequency operation 
is a useful feature of the method for instance in the calculation of 
the intensity of non-resonant Raman scattering,
and in other situations when the target 
frequency range is small but arbitrarily-placed.

In this paper, after a description of the iterative TDDFT method in section
\ref{s:methods}, we apply the method to find the optical response of silver
clusters and shells in section \ref{s:results}. 
We show that the surface-plasmon resonance of atomically-thin shells
is substantially different from the response of compact clusters,
although the difference quickly becomes marginal as the thickness of
the shell is increased. A comparison with other theories and experiment is done 
in section \ref{s:discussion} and section \ref{s:conclusion}
summarizes our results.

\section{Methods\label{s:methods}}

We focus on an \textit{ab-initio} atomistic description of small silver NPs with
a diameter less than 3 nm. The NPs of that size contain typically several
hundreds of atoms which represents a major computational difficulty
for the quantum mechanical description of such systems with most 
current methodologies.
To the best of our knowledge, only TDDFT can currently
cope with such large systems practically within an 
\textit{ab-initio} framework. Moreover, 
due to the size of these systems, we need 
a method of low computational complexity
for which one can envision several candidates. One widely known method
is the wave-packet propagation 
\cite{tddft-prop-dsp:2002,borisov:2013,
octopus-ag-au:2013,jp-prop:2014,gpaw-prop:2015}.
If properly implemented the wave-packet propagation can be realized in 
$O(N)$ operations where $N$ is the number of atoms. A second type of methods 
that are up to the task would be the recently developed stochastic methods
\cite{Neuhauser:2014,Baer:2015} for which a lower-than-linear
computational complexity scaling has been claimed.
A Sternheimer approach to the linear-response
TDDFT seems also a viable alternative \cite{Andrade:2007,Guistino:2014}.
 
The method that we are utilizing in this work is an efficient iterative way 
of solving linear response equations in which we exploit the locality 
of the operators and use an LCAO expansion of the Kohn-Sham (KS)
eigenstates \cite{iter_method,pssb:2010}.
Although the method has a relatively high asymptotic scaling of the 
computational complexity $O(N^3)$, it has been demonstrated 
to be a useful alternative 
to other methods \cite{tddft_iter_plasmons_na380,plasmons_tunable}. 
Moreover, because of the description offered by LCAO, we can
adapt population analysis tools that 
allow connecting the electronic structure of the system
to the chemical intuition (e.g., Mulliken population \cite{coop-book})
to the realm of the optical response as we will describe below.

\subsection{Response functions formalism}

The basic quantity of the linear response TDDFT is the  
density response function $\chi(\bm{r}, \bm{r}',\omega)$, which
is a kernel of an integral operator 
delivering the density change $\delta n(\bm{r},\omega)$ in response 
to a small external perturbation $\delta V_{\mathrm{ext}}(\bm{r},\omega)$

\begin{equation}
\delta n(\bm{r},\omega) = 
\int \chi(\bm{r},\bm{r}',\omega) \delta V_{\mathrm{ext}}(\bm{r}',\omega) dr'.
\label{rf_int}
\end{equation}
In general, the density response function can be expressed
in terms of eigenstates and eigenvalues of the Schr\"odinger equation,
starting the derivation from a perturbative ansatz for the density change
$\delta n(\bm{r},t) = \Psi(\bm{r},t)\delta \Psi(\bm{r},t) + c.c.$
In the case of the KS formulation of DFT, the density response
function becomes more complicated due to the dependence of the KS effective
potential $V_{\mathrm{eff}}(\bm{r}) = 
V_{\mathrm{H}}(n, \bm{r}) + V_{\mathrm{\mathrm{xc}}}(n, \bm{r})$
on the electronic density $n(\bm{r})$ via Hartree and
exchange-correlation (xc) potentials, respectively.
However, the so-called non-interacting response function 
$\chi_0(\bm{r}, \bm{r}',\omega)\equiv 
\frac{\delta n(\bm{r},\omega)}{\delta V_{\mathrm{eff}}(\bm{r}',\omega)}$
remains completely analogue to the Schr\"o{}dinger's response function.
Although the explicit expression of the non-interacting response function
is widely known in the literature \cite{primer-dft,tddft-book},
we will repeat it here for the sake of completeness

\begin{equation}
\chi_0(\bm{r},\bm{r}',\omega) =
\sum_{nm} (f_n-f_m) \frac{
\Psi^*_{n}(\bm{r})\Psi_{m}(\bm{r})
\Psi^*_{m}(\bm{r}')\Psi_{n}(\bm{r}')
}{\omega - (E_{m} - E_{n}) + \mathrm{i}\varepsilon}.
\label{rf0}
\end{equation} 
Here, the occupations $f_n$ and energies $E_n$ of KS eigenstates
$\Psi_{n}(\bm{r})$ do appear and $\varepsilon$ 
is an infinitesimal number that accounts for the proper 
causality of the response. A finite value can be given to $\varepsilon$
in which case it can be thought to represent the finite 
lifetime of excitations.
The interacting response function, defined by equation 
(\ref{rf_int}) can be related to the non-interacting response function 
\cite{PGG:1996,Lecture-Gross:2010} via the so-called interaction kernel 
$K(\bm{r},\bm{r}')
\equiv \frac{\delta V_{\mathrm{eff}}(\bm{r})}{\delta n(\bm{r}')}$

\begin{equation}
\chi(\bm{r},\bm{r}',\omega) = 
\chi_0(\bm{r},\bm{r}',\omega) + 
\int \chi_0(\bm{r},\bm{r}'',\omega) K(\bm{r}'',\bm{r}''')
\chi(\bm{r}''',\bm{r}',\omega) dr'' dr'''.
\label{pgg}
\end{equation}
The TDDFT interaction kernel $K(\bm{r},\bm{r}')$ is commonly separated into 
the Hartree and xc kernels \cite{PGG:1996,Lecture-Gross:2010}
\begin{equation}
K(\bm{r},\bm{r}') = \frac{1}{|\bm{r}-\bm{r}'|} + 
\frac{\delta v_{\mathrm{xc}}(\bm{r})}{\delta n(\bm{r}')}.
\end{equation}
In this work, we mainly use the generalized gradients approximation (GGA)
kernel (see \ref{a:gga-kernel}). The effect of the TDDFT kernel and 
the DFT potential on the optical response is evaluated below in subsection
\ref{ss:dft-funct} by comparing the GGA results with those 
of the local density approximation (LDA).

Because we are interested in optical perturbations the wave-length of 
which exceeds 150 nm, which is much larger than the characteristic length
of our systems, the coupling to the external electromagnetic stimuli 
can be correctly taken into account via a simple dipole operator
$\delta V_{\mathrm{ext}}(\bm{r},\omega) \propto \bm{r}$. 
Furthermore, the far-field
response is connected to the polarizability tensor of the quantum system
\begin{equation}
P_{ij}(\omega) = \int \bm{r}_i \chi(\bm{r},\bm{r}',\omega) \bm{r}_j' dr dr'
\label{Pij-rs}
\end{equation}
which gives rise to the orientation-averaged optical cross section
\cite{Onida-Reining-Rubio:2002}
\begin{equation}
\sigma(\omega) = \frac{4\pi\omega}{3c}  \sum_i P_{ii}(\omega).
\label{cs}
\end{equation}
In this paper we are concerned with the calculation of the 
optical cross section (\ref{cs}) using LDA and GGA DFT functionals and 
a basis set of local orbitals to expand the KS orbitals $\Psi_{m}(\bm{r})$.

\subsection{Product basis set}

The eigenstates entering the response function (\ref{rf0}) are sought
within LCAO 
\begin{equation}
\Psi_{n}(\bm{r}) = \sum_a X^n_a f^a(\bm{r}),
\label{lcao}
\end{equation}
where the expansion coefficients $X^n_a$ are determined in 
a diagonalization procedure. The atomic orbitals $f^a(\bm{r})$ are 
given by a product of radial functions and spherical harmonics.
The LCAO solution is setup and solved within the DFT package 
\textsc{SIESTA} \cite{siesta:2002,siesta:2008}.

The products of eigenstates $\Psi^*_{n}(\bm{r})\Psi_{n}(\bm{r})$
in equation (\ref{rf0}) give rise to products of atomic orbitals 
$f^{a*}(\bm{r})f^b(\bm{r})$. Furthermore, we aim at solving the
integral equation (\ref{pgg})
for the interacting response function. In order to turn this equation 
into an algebraic equation that is easily solved, 
one has to use a \textit{basis set of functions} that are capable to
span the space of atomic-orbital products. This set of basis functions
should be as small as possible
and contain preferably localized functions.
There are several options to construct
such set of functions, hereafter \textit{product basis}. 
The most widely known is probably the auxiliary functions 
for Gaussian basis sets \cite{auxiliary-gauss:1997,auxiliary-demon:2001}. 
However, the SIESTA method is based on 
so-called numerical atomic orbitals (NAO), that can be more flexible and 
economic than Gaussian basis sets. 
There are methods to construct the product 
basis sets for numerical orbitals \cite{ADF:2001,Blase:2004} 
and also our method 
of so-called dominant products \cite{Foerster:2008,df-pk:2009}. 
The dominant product basis can be very accurate but requires
a large number of functions that becomes prohibitive for large systems.
Thus, in this work, we project our dominant products onto a basis 
set of \textit{atom-centered} functions in order to 
reduce the basis set size. As we will see below, the use 
of this more economical basis increases the range of applicability 
and efficiency of the iterative scheme without a significant loss
of accuracy \footnote{We performed many test calculations of 
the optical polarizability comparing the atom-centered functions and 
the dominant product basis sets. For DZP and larger LCAO basis sets, the
discrepancies are negligible and also decrease with system size. For smaller 
basis sets, containing only $s$ and $p$ angular momentum symmetries,
the description worsens, but can be easily recovered by adding 
the higher angular momentum orbitals while generating the atom-centered
product basis.}.

In the method of dominant products, we utilize 
a simple ansatz for the products of atomic orbitals
\begin{equation}
f^{a}(\bm{r})f^b(\bm{r}) = V^{ab}_{\mu} F^{\mu}(\bm{r}).
\label{dpa}
\end{equation}
In this equation, the complex conjugation does not appear
because we use real-valued atomic orbitals \cite{rsh:1997}. 
The product ``vertex'' coefficients $V^{ab}_{\mu}$ and 
the dominant products 
$F^{\mu}(\bm{r})$ are determined in a diagonalization-based procedure.
Namely, we aim at identifying linear combinations of the original
atomic-orbital products $f^{a}(\bm{r})f^b(\bm{r})$ 
\begin{equation}
F^{\mu}(\bm{r}) = \Lambda^{\mu}_{ab} f^{a}(\bm{r})f^{b}(\bm{r}).
\label{df-constr}
\end{equation}
that are 
orthogonal to each other with respect to a Coulomb metric
\begin{equation}
g^{ab, a'b'} = \int f^{a}(\bm{r})f^{b}(\bm{r}) \frac{1}{|\bm{r}-\bm{r}'|}
f^{a'}(\bm{r}')f^{b'}(\bm{r}') dr\, dr'.
\label{cm}
\end{equation}
The linear combinations are built with eigenvectors of the metric
$g^{ab, a'b'}$ 
\begin{equation}
g^{ab, a'b'} \Lambda^{\mu}_{a'b'} = \lambda^{\mu} \Lambda^{\mu}_{ab}
\end{equation}
that guarantee the orthogonality requirement.
Moreover, the eigenvalues of Coulomb metric $\lambda^{\mu}$ are used 
as an indicator of importance of a particular linear combination $\mu$
to the completeness of the basis $\{F^{\mu}(\bm{r})\}$.
Namely, the norm of dominant products $F^{\mu}(\bm{r})$ (\ref{df-constr}),
is proportional to the eigenvalue $\lambda^{\mu}$. Therefore, we 
can consistently limit the number of dominant product (\ref{df-constr})
by ignoring the eigenvector such that
$\lambda^{\mu}$ is lower than a certain eigenvalue threshold.

For our purpose here, it is only necessary to add that 
the procedure is applied to each atom-pair individually.
This keeps the operation count at $O(N)$ scaling, generates localized
dominant products $F^{\mu}(\bm{r})$ and also determines the 
sparsity properties 
of the product vertex coefficients $V^{ab}_{\mu} = \Lambda^{\mu}_{ab}$
in equation (\ref{dpa}). Namely, the vertex coefficients $V^{ab}_{\mu}$
form a \textit{double-sparse table}, which needs asymptotically 
only $O(N)$ stored 
numbers. The term ``double-sparse table'' means that only summation over
two indices of this table generates a full object (vector), while 
the summation
over one of the indices generates a sparse matrix. For instance, 
a summation over the product index $\mu$ generates a matrix 
$s^{ab} = \sum_{\mu} V^{ab}_{\mu}$ 
which has the sparsity of the usual overlap matrix 
$S^{ab}=\int f^a(\bm{r}) f^b(\bm{r}) dr$, while the summation over one 
of the orbitals generates a rectangular sparse matrix 
$R^{a}_{\mu} = \sum_{b} V^{ab}_{\mu}$ because, by construction, a 
dominant product index $\mu$ is connected to orbital indices $a$ and $b$ of
one atom (local pairs) or two atoms with overlapping orbitals
(bilocal pair) rather than to all orbital indices in the molecule.

The dominant products described above have been used in TDDFT, Hedin's $GW$ 
approximation and for solving a Bethe-Salpeter equation 
\cite{iter_method,pssb:2010,GW_small_mol,bse_1}.
However, the construction of dominant products, although mathematically
rigorous and sparsity-preserving, 
has the important disadvantage of generating a large number of functions. 
This disadvantage stems from the construction procedure which is repeated 
independently for each atom pair.
It is easy to see that the dominant products  $F^{\mu}(\bm{r})$
can strongly overlap because different atom pairs can have 
the same or close centers at which the products have their maximal values.
This fact results in a redundant description of the orbital products 
by the dominant product basis when looking from a perspective of the whole 
system. In order to correct for this, we use an ansatz for the auxiliary
basis set that is widely known in quantum chemistry,
\cite{auxiliary-gauss:1997,auxiliary-demon:2001} and also 
in more ``physics-oriented'' proposals \cite{ADF:2001, Blase:2004, aims:2012}. 
Namely, the cited works assume that solely \textit{atom-centered} product 
functions $A^{\mu}(\bm{r})$ are sufficient in practice to express
all orbital products $f^{a}(\bm{r})f^{b}(\bm{r})$.
This very statement, although only justified \textit{a posteriori}, is a useful 
piece of advice that allows to reduce the linear dependencies 
in a product basis set because atom centers are separated
from each other at least with a bonding distance which prevents strong
overlaps of the resulting functions. 

Here we take the local dominant products $F^{\mu}(\bm{r})$ 
(i.e. dominant products generated for orbitals in the same atom,
as opposed to bilocal pairs)
as the atom-centered functions $A^{\mu}(\bm{r})$.
The ansatz for atomic orbitals, analogous to equation (\ref{dpa}) can be 
immediately written as
\begin{equation}
f^{a}(\bm{r})f^b(\bm{r}) = P^{ab}_{\mu} A^{\mu}(\bm{r}),
\label{pba}
\end{equation}
where the atom-centered product vertex coefficients $P^{ab}_{\mu}$
must be still determined. In this work, we extend the ansatz (\ref{pba})
with a recipe for choosing these vertex coefficients $P^{ab}_{\mu}$. 
We propose to draw a sphere around a given atom pair with a radius 
corresponding to the maximal
spatial extension of its orbital products (here defined approximately
as the maximum radius of their atomic orbitals), and consider 
the atom centers within that sphere as \textit{contributing} to 
the atom pair. The ansatz (\ref{pba}) can be easily resolved to obtain 
$P^{ab}_{\nu} = T^{ab,\mu} (v^{\mu\nu})^{-1}$,
where $v^{\mu\nu}$  is the Coulomb matrix element between 
functions $A^{\mu}(\bm{r})$ and $A^{\nu}(\bm{r})$, and 
$T^{ab,\mu}$ is the corresponding matrix element between 
$A^{\mu}(\bm{r})$ and the product of orbitals $f^a(\bm{r})f^b(\bm{r})$.
Notice that in order to termine $P^{ab}_{\nu}$ we do not need 
to invert the whole matrix of Coulomb metric, but a smaller sub-matrix
corresponding to atom-centered functions \textit{inside} the 
contributing sphere. Thus, this step is not computationally prohibitive. 
However, the table $P^{ab}_{\mu}$ has a limited value for 
practical calculations.
Namely, the table $P^{ab}_{\mu}$ can have an order of magnitude more non-zero 
elements that the product vertex coefficients $V^{ab}_{\mu}$. This dramatic 
difference arises because of the distant bilocal atom pairs for which
we have very few dominant products $F^{\mu}(\bm{r})$ instead of many 
atom-centered functions $A^{\mu}(\bm{r})$ contributing to such pairs.

A more fruitful idea proves to be a re-expression of the bilocal 
dominant products $F^{\mu}(\bm{r})$ in terms of atom-centered products 
$ A^{\nu}(\bm{r})$ (that are also chosen with the sphere of contributing
centers)
\begin{equation}
F^{\mu}(\bm{r}) = c^{\mu}_{\nu} A^{\nu}(\bm{r}), 
\label{rea}
\end{equation}
where the projection coefficients $c^{\mu}_{\nu}$ can be also readily
expressed as
\begin{equation}
c^{\mu}_{\nu} = M^{\mu\nu'} (v^{\nu'\nu})^{-1}
\end{equation}
in terms of matrix elements of the Coulomb interaction
\begin{equation}
M^{\mu\nu} = \int \frac{F^{\mu}(\bm{r})A^{\nu}(\bm{r}') }{|\bm{r}-\bm{r}'|} dr\,dr',\,
v^{\mu\nu} = \int \frac{A^{\mu}(\bm{r})A^{\nu}(\bm{r}')}{|\bm{r}-\bm{r}'|} dr\,dr'.
\end{equation}
The projection
ansatz (\ref{rea}) is useful because it is computationally 
very fast to turn a vector in the atom-centered basis to
the dominant product basis 
and back, depending on the quantity that needs to be treated in the product basis.
Namely, we found that it is faster to apply the non-interacting response in the basis of
dominant products and, on the other hand, it is faster to compute and easier to store 
the TDDFT kernel in the basis of atom-centered functions.

\subsection{Iterative method for computing the polarizability}

By inserting the ansatzes (\ref{lcao}) and (\ref{dpa}) into the non-interacting 
response function (\ref{rf0}) we obtain

\begin{equation}
\chi_0(\bm{r},\bm{r}',\omega) = F^{\mu}(\bm{r}) \chi^0_{\mu\nu}(\omega) F^{\nu}(\bm{r}'),
\label{rf0a}
\end{equation}
where the tensor of non-interacting response is given by 

\begin{equation}
\chi^0_{\mu\nu}(\omega) = \sum_{nm} (f_n - f_m)
\frac{(X^n_a V^{ab}_{\mu}X^m_b) (X^m_c V^{cd}_{\nu}X^n_d) }{
\omega - (E_m - E_n) + \mathrm{i}\varepsilon}.
\label{rf0dp}
\end{equation}
Furthermore, we assume for the interacting response function (\ref{pgg})
an expression similar to equation (\ref{rf0a}). This turns 
the Petersilka-Gossmann-Gross equation (\ref{pgg}) into a matrix equation
\begin{equation}
\chi = \chi^0 + \chi^0 K\chi,
\label{pgg-m}
\end{equation}
where the product indices $\mu\nu$ are dropped and the kernel matrix $K$
is given by 

\begin{equation}
K^{\mu\nu} = 
\int \frac{F^{\mu}(\bm{r})F^{\nu}(\bm{r}')}{|\bm{r}-\bm{r}'|} dr\,dr'+
\int F^{\mu}(\bm{r})K_{\mathrm{\mathrm{xc}}}(\bm{r}) F^{\nu}(\bm{r}) dr.
\end{equation}
In this work, we use LDA and GGA kernels 
$K_{\mathrm{\mathrm{xc}}}(\bm{r})=\frac{\delta^2 E_{\mathrm{\mathrm{xc}}}}{
\delta n(\bm{r})\delta n(\bm{r})}$ which are computed in the 
atom-centered functions $A^{\mu}(\bm{r})$.
Explicit expressions for the GGA kernel are
discussed in the \ref{a:gga-kernel}.

The whole response matrix $\chi_{\mu\nu}(\omega)$ is superfluous in the computation 
of the electronic polarizability (\ref{Pij-rs}). Therefore, we further introduce 
the product basis set into the equation for polarizability 
(\ref{Pij-rs}) and using equation (\ref{pgg-m}) obtain 
\begin{equation}
P_{ij}(\omega) = d^{\mu}_i (\delta^{\mu}_{\nu'} - \chi^0_{\mu\nu'}(\omega)K^{\nu'\mu'})^{-1}
\chi^0_{\mu'\nu}(\omega) d^{\nu}_j,
\label{Pij-pb}
\end{equation}
where the dipole moments of the product functions 
$d^{\mu}_i = \int F^{\mu}(\bm{r})\bm{r}_i\, dr$ appear.
The calculation of the polarizability 
(\ref{Pij-pb}) is split into a calculation of the density change
in the product basis $\delta n_{\mu}(\omega)$
\begin{equation}
[\delta - \chi^0(\omega) K] \delta n(\omega) = \chi^0(\omega) d \label{dn-pb}
\end{equation}
and a final trace with the dipole moments $P_{ij}(\omega)= 
d^{\mu}_i \delta n_{\mu,j}(\omega)$. The linear equation (\ref{dn-pb})
is solved with an iterative method similar to the Arnoldi method and optimized
for delivering the polarizability $P_{ij}(\omega)$ with a given precision
rather than the density change $\delta n_{\mu,j}(\omega)$ \cite{iter_method}
that would be the target for general-purpose iterative solvers \cite{Luc:2005}.

Iterative linear equation solvers require only the action of a 
matrix onto given vectors to find the solution $\delta n_{\mu,j}(\omega)$.
In our case, the matrix reads $A=[\delta - \chi^0(\omega) K]$. The product
of this matrix with a vector $z$ can be computed in terms of subsequent 
matrix-vector products of the TDDFT kernel $K$ with a vector and of the 
non-interacting response function $\chi^0$ with another vector. The former 
product is easy to organize because we represent the kernel as
a full matrix between the atom-centered product functions. The latter
product is more involved and is explained below.

The matrix-vector product of the non-interacting response function 
$\chi^0_{\mu\nu}(\omega)$ with a vector $z^{\nu}$ is also split 
into a sequence matrix-vector and matrix-matrix operations. First,
we use the projection ansatz (\ref{rea}), insert that expression into 
the equation (\ref{rf0a}) and get

\begin{equation}
\chi_0(\bm{r},\bm{r}',\omega) = 
A^{\mu}(\bm{r}) c^{\tilde\mu}_{\mu}
\chi^0_{\tilde\mu\tilde\nu}(\omega)
c^{\tilde\nu}_{\nu} A^{\nu}(\bm{r}'),
\label{rf0ac}
\end{equation}
where the tilde indices run over the dominant products and the 
simple indices run over the atom-centered product functions. 
Equations (\ref{rf0dp}) and (\ref{rf0ac})  define the matrix 
expression for the matrix-vector product $\chi^0_{\mu\nu}(\omega) z^{\nu}$.
There are several sequences of operations 
to organize the matrix-vector product $\chi^0_{\mu\nu}(\omega) z^{\nu}$.
However, our original algorithm suggested in 
\cite{iter_method} proved to be the fastest alternative, with 
moderate memory requirements. The algorithm starts with precomputing
of a quantity $\alpha^{an}_{\tilde\mu}=V^{ab}_{\tilde\mu} X^n_b$ 
for occupied eigenstates states $n$. The table  $\alpha^{an}_{\tilde\mu}$ is
stored in a block-sparse storage that uses $O(N^2)$ elements of the random
access memory (RAM) and enables matrix operations with ordinary
basic linear algebra subroutines \cite{netlib-blas}. The table $\alpha^{an}_{\tilde\mu}$
is the major ``memory consumer'' which, however, can be obviously eliminated
when larger systems need to be treated. A vector $z^{\mu}$, on which
the response function has to be applied, is converted to the basis of 
dominant products $z^{\tilde\mu}=c^{\tilde\mu}_{\mu}z^{\mu}$, then 
the product indices are summed to produce a matrix 
$\beta^{an}=\alpha^{an}_{\tilde\mu}z^{\tilde\mu}$. The matrix 
$\beta^{an}$ is a full rectangular matrix. The calculation
continues with a matrix-matrix multiplication 
$\gamma^{mn} = X^m_a\beta^{an}$,
where the index $m$ now runs over the unoccupied KS orbitals.
The latter multiplication determines
a maximal computational complexity of the whole algorithm which is 
$O(N^3)$. The calculation continues with an update of the matrix
$\gamma^{mn}$ with the frequency-occupation mask 
$\tilde\gamma^{mn}=\gamma^{mn}\left(
(f_n-f_m)/(\omega - (E_m-E_n) + \mathrm{i}\varepsilon)-
(f_n-f_m)/(\omega + (E_m-E_n) + \mathrm{i}\varepsilon)\right)$.
The next $O(N^3)$ matrix-matrix multiplication reads 
$\tilde\beta^{an} = X^m_a\tilde\gamma^{mn}$. Finally the non-interacting 
density change $\delta n^0_{\tilde\mu}=\chi^0_{\tilde\mu\tilde\nu}z^{\tilde\nu}$
is obtained by tracing over $a$ and $n$ indices in the product of 
$\tilde\beta^{an}$ with the precomputed quantity $\alpha^{an}_{\tilde\mu}$,
i.e. $\delta n^0_{\tilde\mu} = \tilde\beta^{an}\alpha^{an}_{\tilde\mu}$.

In summary, we described the iterative algorithm of $O(N^3)$
computational complexity that uses $O(N^2)$ memory and enables
a relatively fast calculation of interacting polarizability
in plasmonic systems, i.e. in systems that have many nearly-degenerate
transitions. Although the presented algorithm possesses a relatively high 
asymptotic computational complexity, the algorithm is relatively 
inexpensive in terms of computational resources. 
This allowed us performing calculations for system sizes
containing hundreds of atoms, despite the fact that our implementation
uses only OpenMP parallelization. For example, 
the calculation of Ag$_{561}$ icosahedral cluster has been done on 
a 12 core node (Intel Xeon CPU X5550\@ 2.67GHz, release date 2009)
in 25 hours of walltime. The iterative procedure took most of the walltime 
(19.7 hours) while performed for 200 frequencies and only for 
xx- component of polarizability tensor (\ref{Pij-pb}).

\subsection{Accuracy of the methods \label{ss:acc}}

In the algorithm presented above, several approximations are involved.
The approximations originating from the input DFT calculation,
including the choice of the xc functional and 
the applied basis set of atomic orbitals, are discussed below 
in subsections
\ref{ss:dft-funct} and \ref{ss:ao-basis}. 
The approximations originating from the implementation of TDDFT 
are related to the usage of the product basis sets and 
an iterative procedure to compute the induced density change
for a given perturbation.
Both of the latter approximations have been carefully analysed
in our the previous works in which we identified corresponding accuracy
indicators. The accuracy indicators include the difference between 
the overlap-
and dipole- matrix elements computed directly from the atomic orbitals
and via the moments of the product basis functions, and 
a convergence test of the iterative procedure. The accuracy indicators 
have been routinely controlled in the calculations we
present in this work. As a result of this control, we found that the 
spectra presented in this work are unaffected by the 
usage of product basis sets until a high frequency approximately 
$\omega=50$~eV for all cluster sizes; the iterative procedure provides 
the same results for polarizability, within a given, previously specified
small tolerance, as computed via Casida formulation 
(possible only for small clusters with less than about 20 atoms 
in our realization).

\subsection{Choice of the exchange-correlation functional \label{ss:dft-funct}}

All correlation effects should be captured in DFT through a single
xc functional. Previous studies revealed that both the geometry 
\cite{Blaha:2009} and the 
electronic structure \cite{yabana:1999,gga-vs-lrc:2014,gpaw-prop:2015} of silver
containing compounds are affected by the choice of xc functional. 
In the extensive comparative study in Ref. \cite{Blaha:2009}, it was shown
that the lattice parameter predicted by LDA is 1.6\% shorter than the 
experimental value. This situation is improved by some GGA functionals.
In particular, the functionals by Wu and Cohen (WC) \cite{wc-funct},
the Perdew-Burke-Ernzerhof adapted to solids \cite{pbesol-funct},
and that due to Armiento and Mattsson \cite{am05-funct} provided
the best results (deviations of the lattice parameter below 0.5\%).
Because the WC functional shows the best performance for bulk silver and 
performs well also for finite systems \cite{wc-better-geom:2007},
we have chosen this functional for geometry optimization.

\begin{figure}[htbp]
\includegraphics[width=14cm]{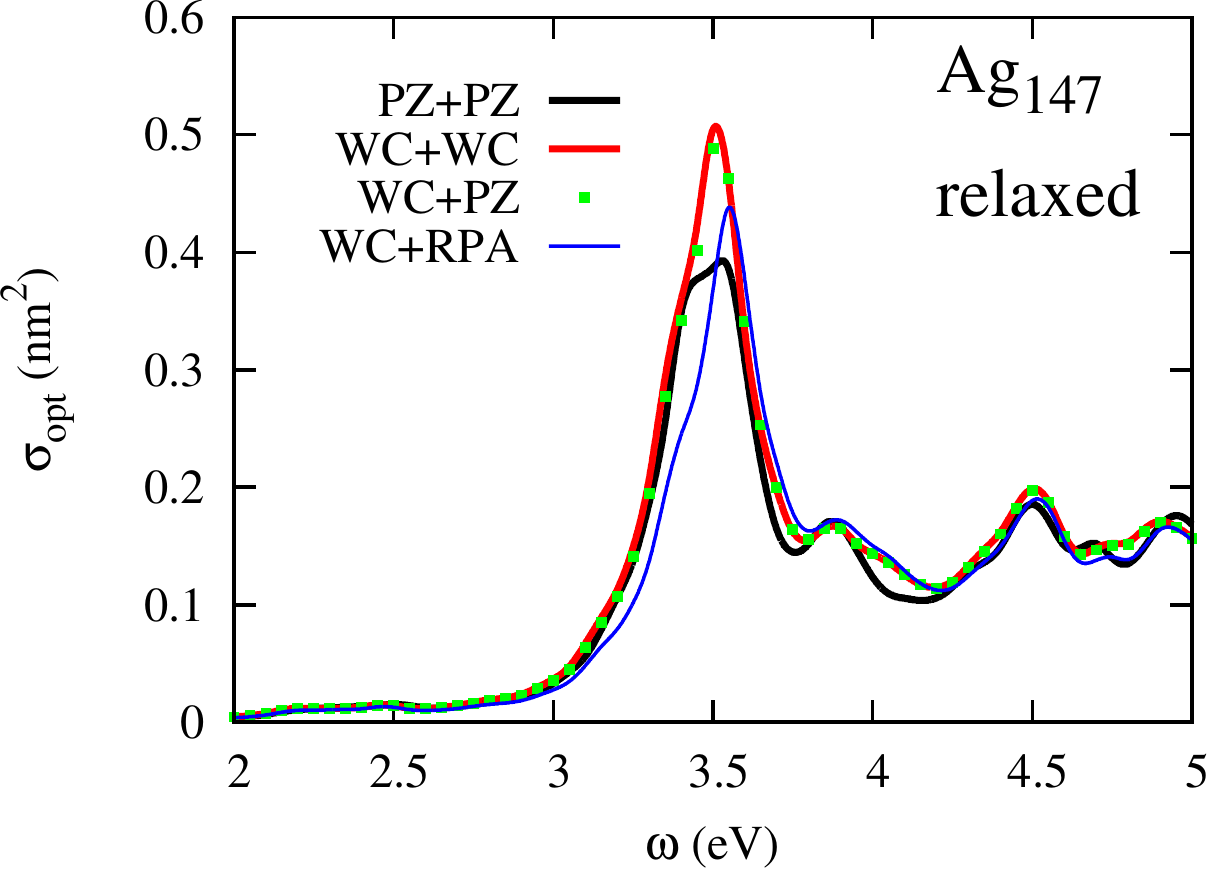}
\caption{\label{f:lda-gga}Absorption cross section of 
the icosahedral Ag$_{147}$ cluster computed with LDA and GGA
functionals. The curves are labeled according to the xc potential
(first label) and xc kernel (second label) used. 
The Perdew-Zunger (PZ) LDA functional and 
Wu-Cohen (WC) GGA functionals are compared.
Blue solid line represent RPA $K_{\mathrm{xc}}=0$ starting from WC
results. Relaxation of geometries was done with PZ or WC
functionals.}
\end{figure}

With respect to electronic structure, it is well known from the literature that
LDA and GGA predict a too low onset of the $d$-bands in the solid and this reduces 
the intensity of the low-frequency plasmonic resonance produced mainly by
the $s$-electrons \cite{Marini-lda-gw:2002}. This deficiency of LDA and GGA
can be corrected by using so-called long-range-corrected (LRC) xc functionals 
which contain a portion of Fock exchange 
\cite{abinitio-cis:2012,small-ag-cu:2014,gga-vs-lrc:2014},
the van-Leeuwen-Baerends explicit ansatz with the correct asymptotic behavior
\cite{Stener-gold:2011}, or orbital-dependent functionals
\cite{Kuisma-GLLB-SC:2010,gpaw-prop:2015}.
Unfortunately, these functionals are not available within the publicly-available
version of the 
SIESTA package \cite{siesta:2002}. Moreover, the LRC functionals 
referenced above give rise to a non-local two-point TDDFT kernel, 
which is either not known or 
computationally too expensive 
to treat the large systems addressed here. From this point of view, 
the Sternheimer approach and wave-packet propagation approach have an 
advantage versus iterative TDDFT: former approaches only need 
the xc potential, and do not involve the TDDFT kernel.
Fortunately, local and semi-local functionals correctly capture trends
of the plasmonic response in nano-particles as was well documented in the past
\cite{yabana:1999}, and are still widely used \cite{weissker-gga-vs-lrc:2015}.

For these reasons, we have the WC functional for both, 
the DFT and TDDFT steps of our calculation. 
In order to assess the effect of the xc functional, 
we compare the LDA and GGA spectra and also analyse the contributions 
from $d$-electrons 
to the total absorption cross section (in section \ref{ss:analysis}
and \ref{ss:analysis-results}). In the figure \ref{f:lda-gga} we 
compare the optical absorption cross-sections computed with LDA and GGA
functionals for the Ag$_{147}$ icosahedral cluster. The geometries of the 
cluster were relaxed with Perdew-Zunger (PZ) LDA or WC functionals, although
the relaxations themselves did not affect the spectra significantly.
We can immediately confirm that GGA increases the intensity of 
the plasmon peak as compared with LDA. However, it does not significantly 
shift the resonant frequencies (3.5 eV LDA, 3.54 GGA).
Moreover, it is interesting to note that the effect of the gradient corrections
in the kernel $K_{\mathrm{xc}}$ is marginal. Namely, in figure \ref{f:lda-gga}
we can hardly distinguish the spectra in which the GGA kernel is substituted 
by the LDA kernel and the other computational parameters are kept the 
same (WC+WC versus WC+PZ curves). As expected, however, the effect of 
Hartree kernel is crucial. The non-interacting response (not shown in the 
plots) is dominated at low energies by a peak at approximately 1 eV, 
way too low as compared to experiment.
Therefore, we see a minor influence of xc kernel 
on the optical response of our Ag nanoparticles.
It is important to note, however, that the improved approximation 
to the quasi-particle spectrum provided by the KS eigenvalues
computed with the GGA functional is reflected
in an improved description of the
optical properties. This is an important information from a methodological
point of view because
the calculation of real-space integrals of the xc kernel is much more time
consuming (two orders of magnitude) than the calculation of the Hartree kernel.
The Hartree kernel is calculated with the help of fast Bessel transforms
\cite{Talman:2009, Talman:2010}
 and the multipole expansions \cite{df-multipole,iter_method}.
Moreover, the GGA xc kernel is way more cumbersome than the LDA kernel
(see \ref{a:gga-kernel}) and for many of the most sophisticated functionals 
the explicit expressions of the kernel are still lacking. 
The analysis presented above with respect to the influence of the 
xc kernel (RPA versus GGA) is consistent to the
previously published for smaller clusters \cite{xc-contr-ag13-ag55:2010}.

\subsection{Choice of atomic orbital basis set\label{ss:ao-basis}}

The choice of the atomic-orbital basis set determines the quality of 
LCAO calculations to a large extend. The numerical orbitals used in 
the SIESTA package are capable to approach the results of plane-wave
calculations 
for bulk systems \cite{nao-vs-pw:2001,siesta-orb:2002}, at least 
for ground state properties and the description of the low lying
unoccupied states. At the same time,
it was found that semi-infinite systems (surface properties) 
need special care when being described using confined NAO 
\cite{siesta-flt-orb:2009}. Namely, it was found that adding 
a single layer of floating orbitals can considerably improve 
the surface properties
with only a small impact on the computational performance.
Because the surface-to-volume ratio of our target clusters is rather
large \cite{Heer:1993}, we assess the quality of the default 
SIESTA basis set by augmenting it with an extra layer 
of floating orbitals.

\begin{figure}[htbp]
\includegraphics[width=14cm]{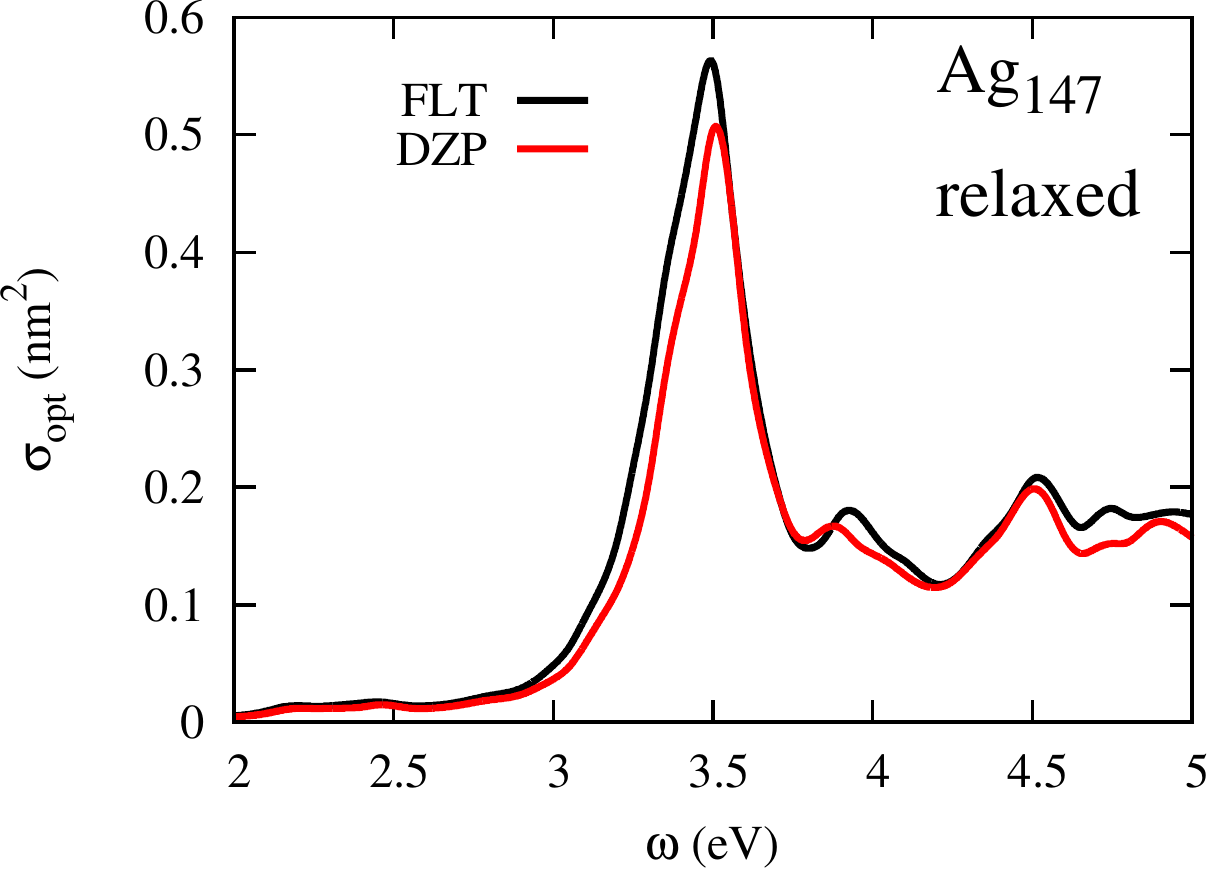}
\caption{\label{f:dzp-flt}Absorption cross section of 
the icosahedral Ag$_{147}$ cluster computed with a DZP basis set 
and with a DZP basis augmented with an extra layer of 
floating orbitals (FLT).
Relaxations have been done for all atoms except 
``ghost atoms'' that carry floating orbitals.}
\end{figure}
 
In figure \ref{f:dzp-flt}, we plot the absorption spectra of the Ag$_{147}$
cluster computed with a double-zeta polarized (DZP) basis (EnergyCutoff=50 meV)
and with the same DZP basis augmented with an extra 
layer of $s$- and $p$- floating orbitals. 
The floating orbitals are placed at the 
positions of the next (fifth) Mackay layer 
(geometry of clusters is discussed below in section \ref{s:results}) and are 
excluded from the relaxation procedure. The WC functional was used 
in both calculations. One can see on the figure
that the extra layer of floating orbitals slightly redshifts the 
frequency of main plasmon resonance (by 0.02 eV) and 
increases the absorption cross section around the resonance.
A direct analysis of the absorption cross section in terms of 
the different cluster layers (see section \ref{ss:analysis})
shows that the enhancement is due to the extra layer of 
``ghost atoms'' which carry the floating orbitals 
(see subsections \ref{ss:analysis} and \ref{ss:analysis-results})
and that the main contribution to the cross section is still 
due to the outer layer of real atoms. 
At the same time, the layer of floating orbitals
does not affect the computation cost dramatically because 
it does not unnecessarily 
decrease the sparsity of the matrices involved in 
the calculation, which is the case when spatially extended
diffuse orbitals are added to the basis set.

\subsection{Analysis of the interacting density change.\label{ss:analysis}}

Here we focus on the analysis of the interacting polarizability 
in terms of the spatial distribution of atoms and in terms of 
the non-interacting electron-hole pairs. Other type of analysis
are too technical for a general presentation and 
we reserved them for \ref{a:analysis}.

\subsubsection{Contribution of different atoms to the polarizability.}

In our framework, the interacting polarizability $\alpha(\omega)$
is given by 
\begin{equation}
\alpha(\omega) = 
d^{\mu}\delta n_{\mu}(\omega), \ 
d^{\mu} = \int F^{\mu}(\bm{r}) \bm{r} dr,
\end{equation}
where we drop all Cartesian indices for the sake of clarity.
The index $\mu$ runs over all product functions  $F^{\mu}(\bm{r})$,
which are centered on real and ghost atoms. 
The sum over product functions can be 
split into sub-sums according to a given criterion. For instance,
we can split the sum over the product indices into 
sub-sums over atomic layers in the cluster $L$:
$\alpha(\omega) = \sum_{L} \alpha_L(\omega)$ with 
a partial polarizability $\alpha_L(\omega)$  given by
\begin{equation}
\alpha_L(\omega) = d^{\mu} P_{\mu}^{\nu}(L) \delta n_{\nu}(\omega).
\label{pol_tags}
\end{equation}
Here a padding operator $P_{\mu}^{\nu}(L)$ is introduced. The padding 
operator is a diagonal matrix whose elements are equal to one
$P_{\mu}^{\mu}(L)=1$ if the product index $\mu$ belongs to an atom
in the layer $L$ and zero otherwise.

\subsubsection{Electron-hole expansion of the 
interacting induced density.\label{ss:ph-mapping}}

The analysis presented above for the interacting polarizability 
can be completed using an expansion of the interacting density change 
$\delta n(\bm{r}, \omega)$ in terms of electron-hole pairs

\begin{equation}
\delta n(\bm{r}, \omega) = \sum_{ij}
\delta n_{ij}(\omega) \Psi^*_i(\bm{r})\Psi_j(\bm{r}),
\label{dn:eh}
\end{equation}
where $\Psi_j(\bm{r})$ are KS eigenstates.

This kind of expansions are naturally arising in the Casida formulation
of TDDFT \cite{Casida:1995} and this is perhaps at the root of the
popularity of Casida's formulation, because the electron-hole 
expansion (\ref{dn:eh}) allows classifying a given excitation 
in terms of the character of the 
non-interacting transitions contributing to it,
as was recently 
elaborated with an alternative method \cite{Ullrich:2015}.
 
Obtaining the expansion (\ref{dn:eh}) is relatively 
straightforward in the iterative formulation of TDDFT 
presented above.
The interacting density change $\delta n(\bm{r},\omega)$
is given by (compare with equation (\ref{rf_int}))

\begin{equation}
\delta n(\bm{r},\omega) = \int \chi_0(\bm{r},\bm{r}',\omega) 
\delta V_{\mathrm{eff}}(\bm{r}',\omega) dr',
\label{rf_int0}
\end{equation}
where $\delta V_{\mathrm{eff}}(\bm{r},\omega)$ is an effective (screened)
perturbation. Using sum-over-states representation (\ref{rf0})
of the non-interacting density response function we can immediately
see the electron-hole expansion coefficients $\delta n_{ij}(\omega)$

\begin{equation}
\delta n_{ij}(\omega) = \frac{ f_i-f_j
}{\omega - (E_{j} - E_{i}) + \mathrm{i}\varepsilon}
\int \Psi^*_{j}(\bm{r}')\Psi_{i}(\bm{r}')
\delta V_{\mathrm{eff}}(\bm{r}',\omega) dr'.
\label{dn_ij}
\end{equation}
Now it remains only to derive an equation for the effective
perturbation $\delta V_{\mathrm{eff}}(\bm{r},\omega)$.
Using the Petersilka-Gossmann-Gross equation (\ref{pgg})
and the two alternative expressions 
(\ref{rf_int0}) and (\ref{rf_int})
for the interacting density change, it is straightforward
to derive an equation for the effective perturbation

\begin{equation}
\left[\delta(\bm{r}-\bm{r}') - K(\bm{r},\bm{r}'') \chi_0(\bm{r}'',\bm{r}',\omega)\right]
\delta V_{\mathrm{eff}}(\bm{r}',\omega) = \delta V_{\mathrm{ext}}(\bm{r},\omega).
\end{equation}

In terms of product basis (\ref{dpa}) and for optical absorption
$\delta V_{\mathrm{ext}}(\bm{r},\omega)\equiv\bm{r}$, 
the latter equation transforms into a linear algebraic equation
\begin{equation}
\left[\delta - K\chi_0(\omega)\right]\delta V_{\mathrm{eff}}(\omega) = d,
\end{equation}
where we dropped Cartesian indices and product indices for clarity.
This equation can be solved iteratively with a generalized minimal
residue solver \cite{Luc:2005}.
Using the product basis again in equation (\ref{dn_ij}) we get
\begin{equation}
\delta n_{ij}(\omega) = \frac{ (f_i-f_j)
(X^{j}_a V^{ab}_{\mu} X^{i}_b)
}{\omega - (E_{j} - E_{i}) + \mathrm{i}\varepsilon}
\delta V^{\mu}_{\mathrm{eff}}(\omega).
\label{dn_ij_pb}
\end{equation}

Having the density change in terms of electron-hole pairs (\ref{dn_ij_pb})
we can define a transition-resolved optical polarizability
\begin{equation}
\alpha_{ij}(\omega) = \delta n_{ij}(\omega) 
\int \Psi^*_{j}(\bm{r}) \bm{r} \Psi_{i}(\bm{r}) dr
\label{p_ij_pb}
\end{equation}
and thus assess the contribution of each non-interacting pair of 
states to the true, interacting polarizability. 
Moreover, it is now possible to answer questions 
related to the symmetry of the charge density with a strongest
contribution at a given frequency \cite{Ag13-lda-analysis:2008}
and perform other types of analysis analogous 
to the crystal orbital overlap and Hamiltonian populations analysis 
of density matrix \cite{coop-book}. Here, we will focus on the 
analysis of the polarizability in terms of the dominant 
atomic angular-momentum contributions in the initial state. 
For this, we will explicitly separate the 
occupied and virtual states and look for the electron-hole
expansion in the form 
\begin{equation}
\delta n(\bm{r}, \omega) = 
\sum_{i\in \mathrm{occ}, j\in\mathrm{unocc}}
\delta n_{ij}(\omega) \Psi^*_j(\bm{r})\Psi_i(\bm{r}),
\end{equation}
where the expansion coefficients $\delta n_{ij}(\omega)$ read

\begin{equation}
\delta n_{ij}(\omega) = \left( \frac{ (f_i-f_j)
(X^{i}_a V^{ab}_{\mu} X^{j}_b)
}{\omega - (E_{j} - E_{i}) + \mathrm{i}\varepsilon}-
\frac{ (f_i-f_j)
(X^{i}_a V^{ab}_{\mu} X^{j}_b)
}{\omega + (E_{j} - E_{i}) + \mathrm{i}\varepsilon} \right)
\delta V^{\mu}_{0}(\omega).
\label{dn_ij_pb_ov}
\end{equation}

Now we can define a polarizability resolved in the angular momentum $l$ of 
the occupied states
\begin{equation}
\alpha_{l}(\omega) = \sum_{i\in \mathrm{occ}, j\in\mathrm{unocc}} 
\delta n_{ij}(\omega) X^j_a d^{ab} \delta_{l_a,l} X^i_a.
\label{pp-occ-virt}
\end{equation}
Obviously, the partial polarizability $\alpha_{l}(\omega)$ add up to 
the total interacting polarizability $\alpha(\omega) = \sum_l \alpha_{l}(\omega)$
and each of the partial polarizabilities gives an idea of the contribution of 
a given symmetry in the occupied states to the total polarizability.
The result of this analysis for silver clusters
is presented below in subsection \ref{ss:analysis-results}.

\section{Results\label{s:results}}

In this work, we focus our attention on silver clusters and shells, 
which are well known plasmonic systems. We will address 
the dependence of the plasmonic resonances on the system size,
the thickness of the shells and the details of the cluster
geometry (relaxation method).

\subsection{Calculation parameters}

Clusters of icosahedral shapes were constructed using the atomic simulation
environment \cite{ase:2002} according to a Mackay motif \cite{mackay:2002}.
The initial atomic positions were obtained using a  4.0 \AA{} 
lattice constant that is close to the GGA-relaxed geometry and is smaller 
than experimental lattice constant for bulk Ag (4.09 \AA{}). 
The ``ideal'' geometries were relaxed by minimizing the forces
acting on atoms below 0.02 eV/\AA{} using the GGA functional 
after Wu and Cohen \cite{wc-better-geom:2007,Blaha:2009}, 
as we discussed in subsection \ref{ss:dft-funct}.
Otherwise said, we used similar parameters in all calculations. The spatial extension 
of orbitals was set in a default procedure by an EnergyShift
parameter of 50 meV and a double-zeta polarized (DZP) set 
of atomic orbitals was used. The real-space mesh has been set 
via a Meshcutoff=150 Ry. Only valence electrons (5s$^1$4d$^{10}$)
are represented in the atomic orbitals set, which give rise 
to 15 atomic orbitals per atom in the DZP basis set.
Moreover, we added a layer of floating orbitals 
(see subsection \ref{ss:ao-basis})
in the cluster calculations and two layers of floating orbitals---inner and
outer---in the case of shell geometries. Coordinates of these ``ghost atoms''
were kept fixed during relaxations.

The core electrons are removed by using the pseudo-potential (PP) of
Troullier-Martins type. The PPs were generated with the ATOM program,
part of the SIESTA distribution.
The parameters for PP generation were taken from the SIESTA database 
\cite{siesta-pp-db} except for 
the use of Wu and Cohen functional. 
It is interesting
to note that the optimized PP described in Ref.~\cite{rivero:2015},
which is supposed to provide a band structure
in better agreement with all-electron calculations, 
failed to satisfactorily describe the optical absorption cross section.
Indeed, the description is severely worsen and the main plasmonic 
resonance disappears.

\subsection{Relaxed geometries}

\begin{figure}[htbp]
\begin{tabular}{p{0.1cm}p{4.5cm}p{4.5cm}p{4.5cm}}
   & \centerline{S3}  & \centerline{S4} & \centerline{S5} \\
L1 &
\includegraphics[width=4.5cm]{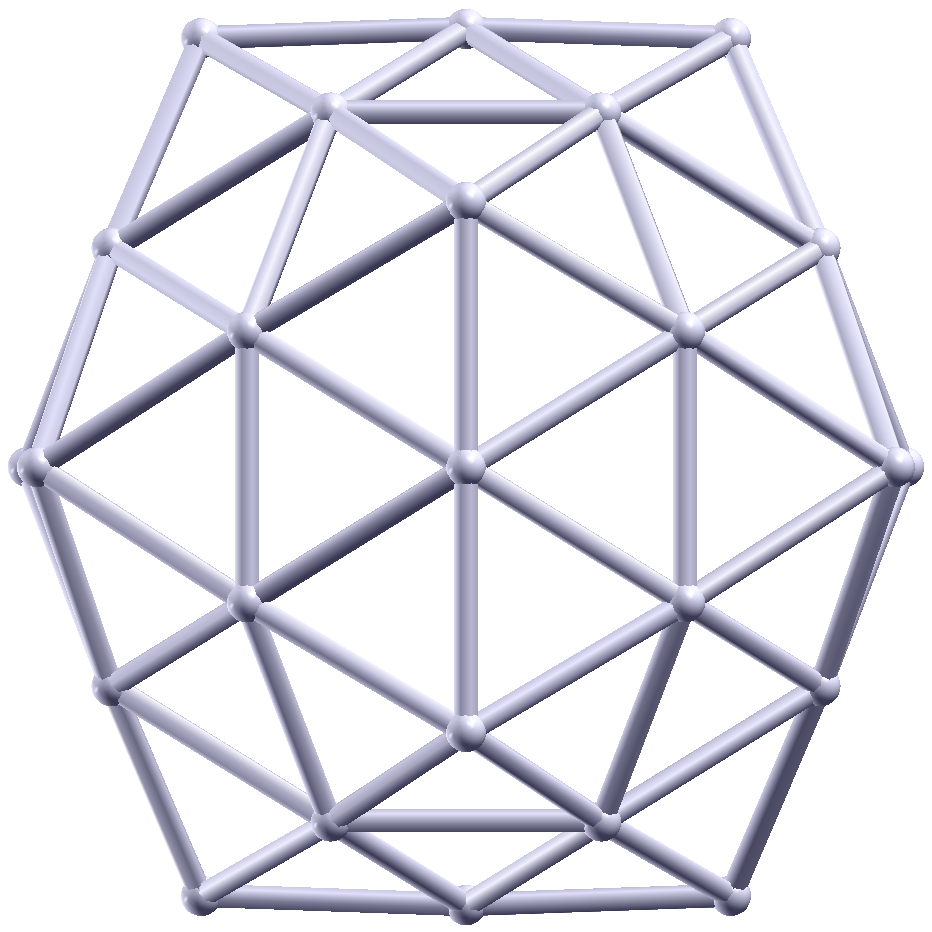} &
\includegraphics[width=4.5cm]{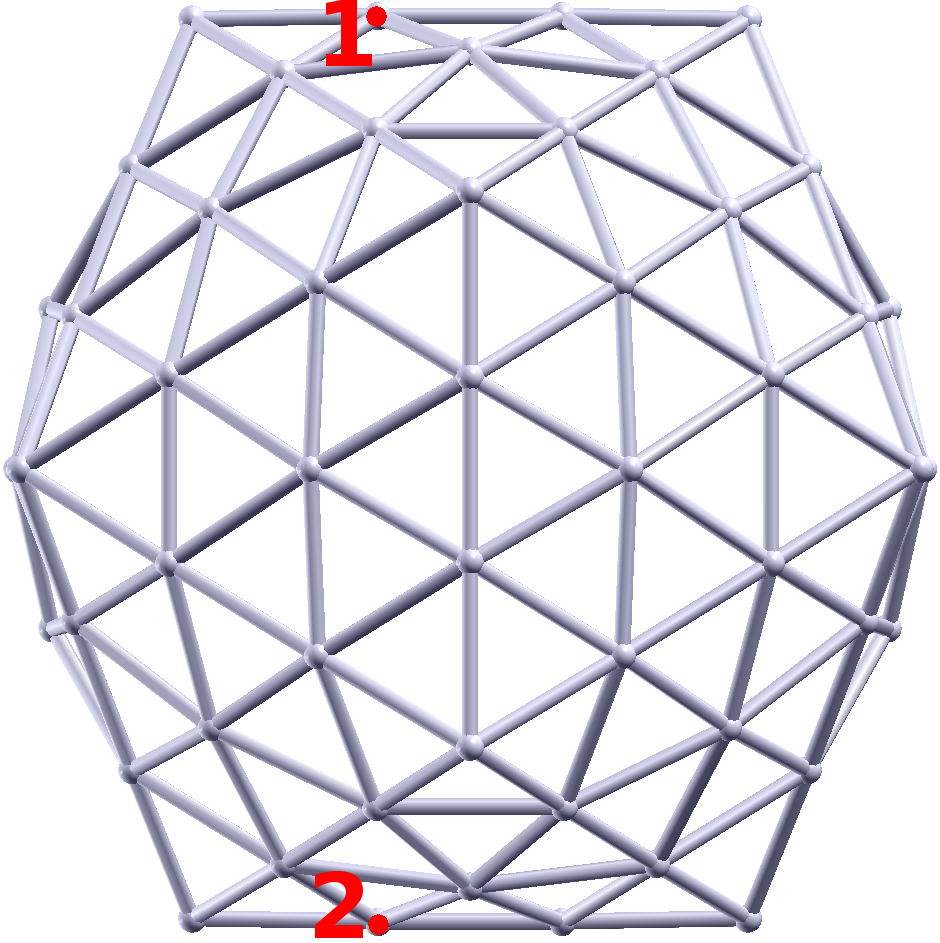} &
\includegraphics[width=4.5cm]{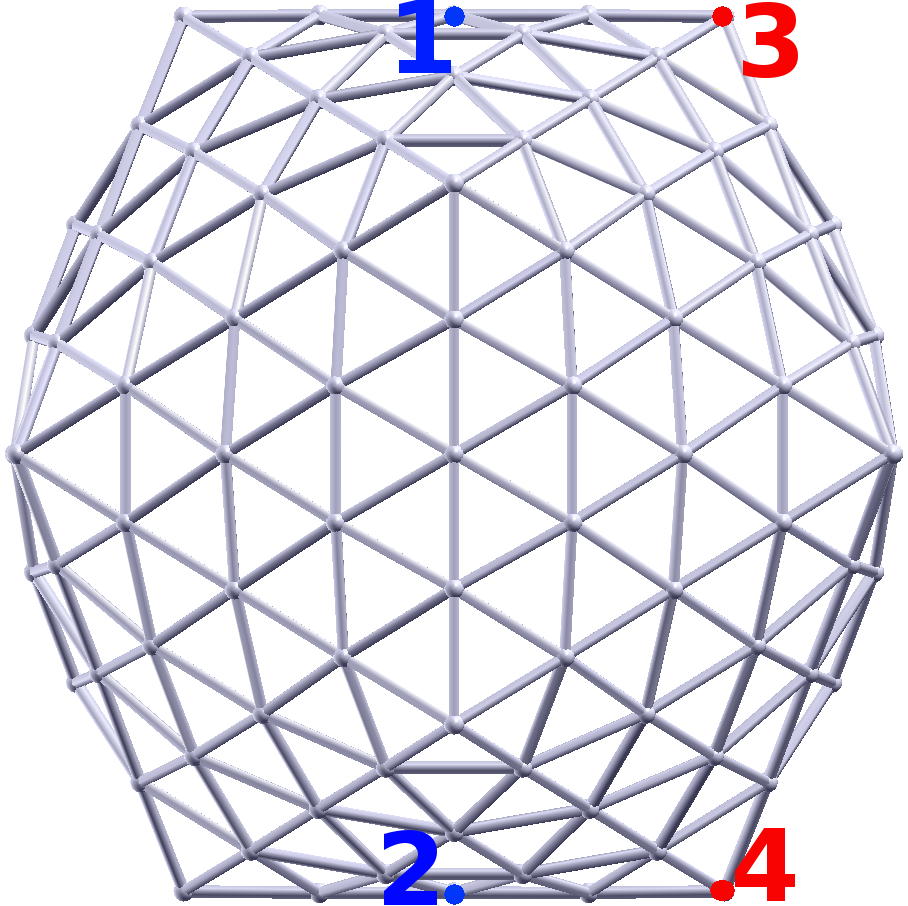} \\
L2 &
\includegraphics[width=4.5cm]{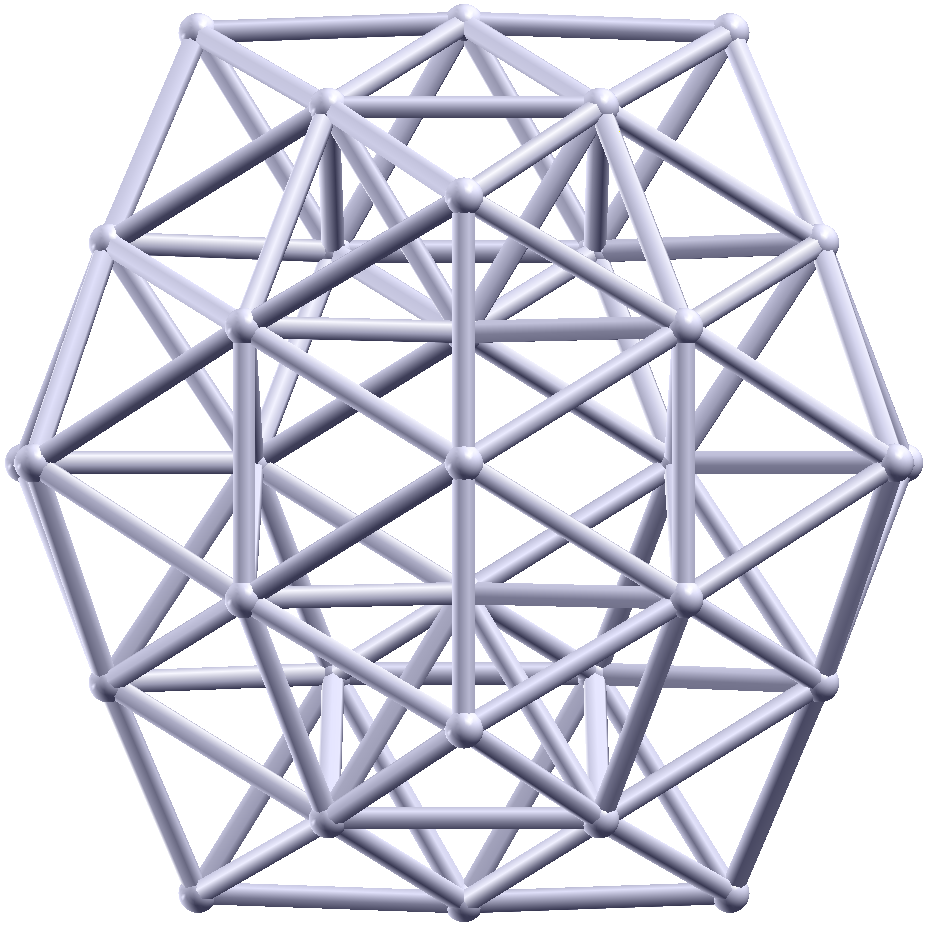} &
\includegraphics[width=4.5cm]{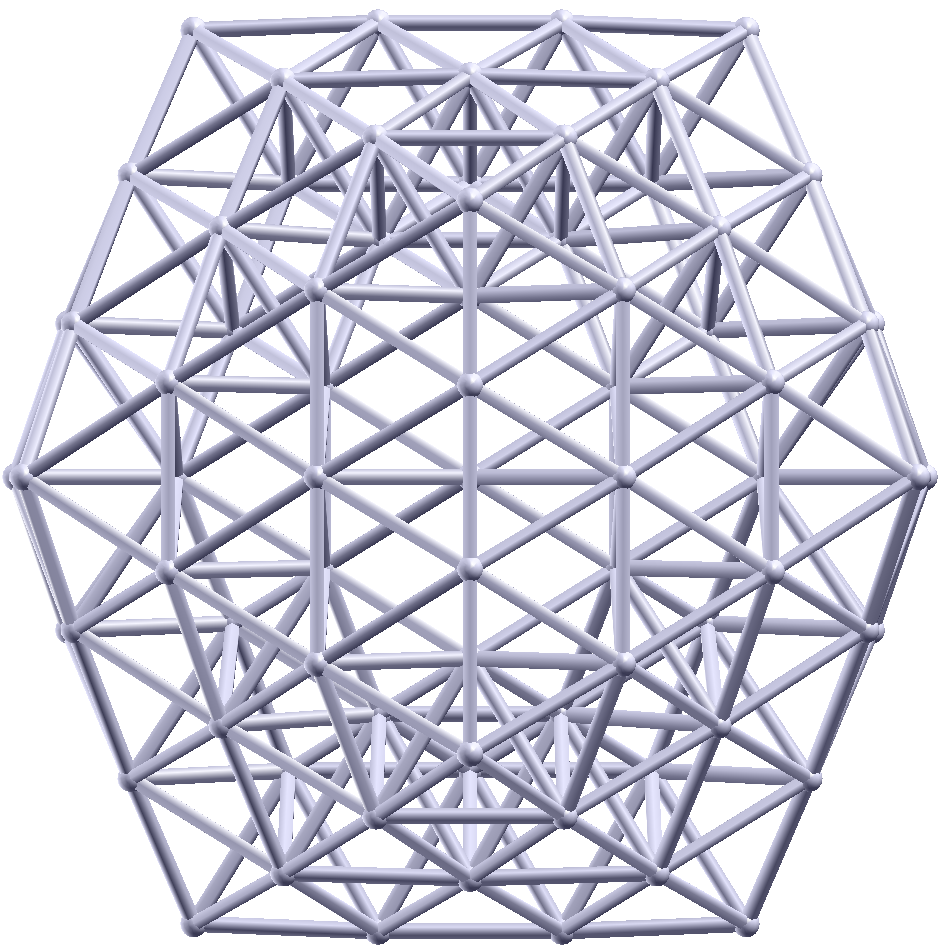} &
\includegraphics[width=4.5cm]{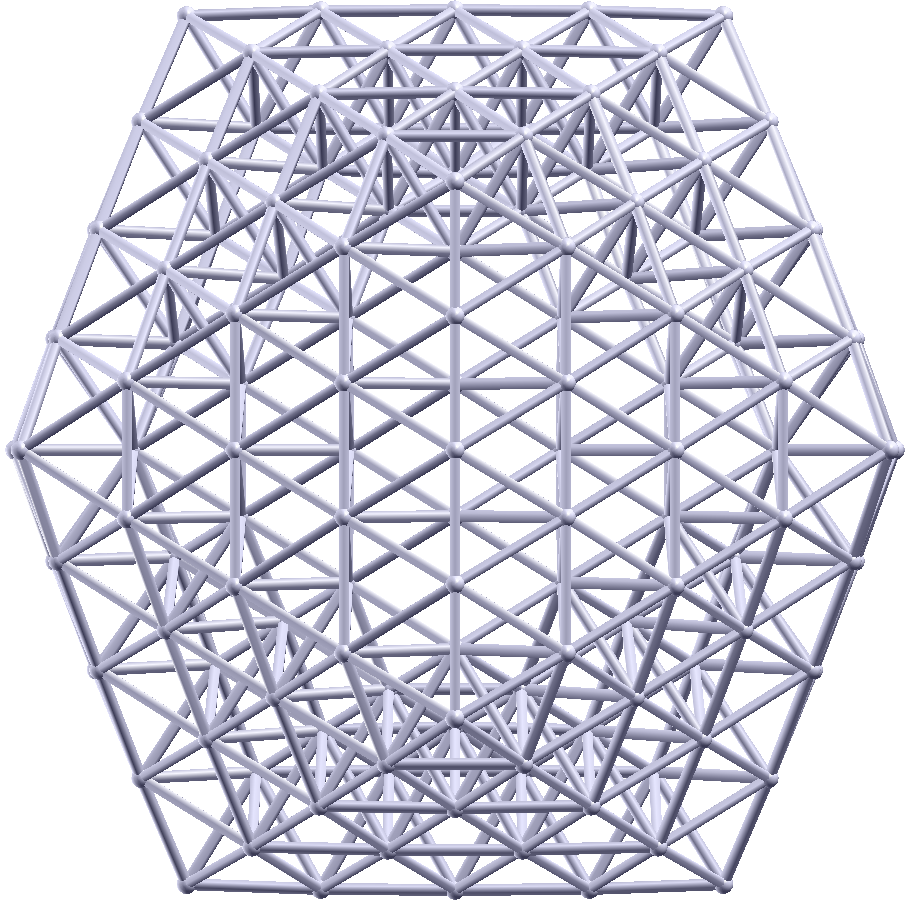} \\
 &
\includegraphics[width=4.5cm]{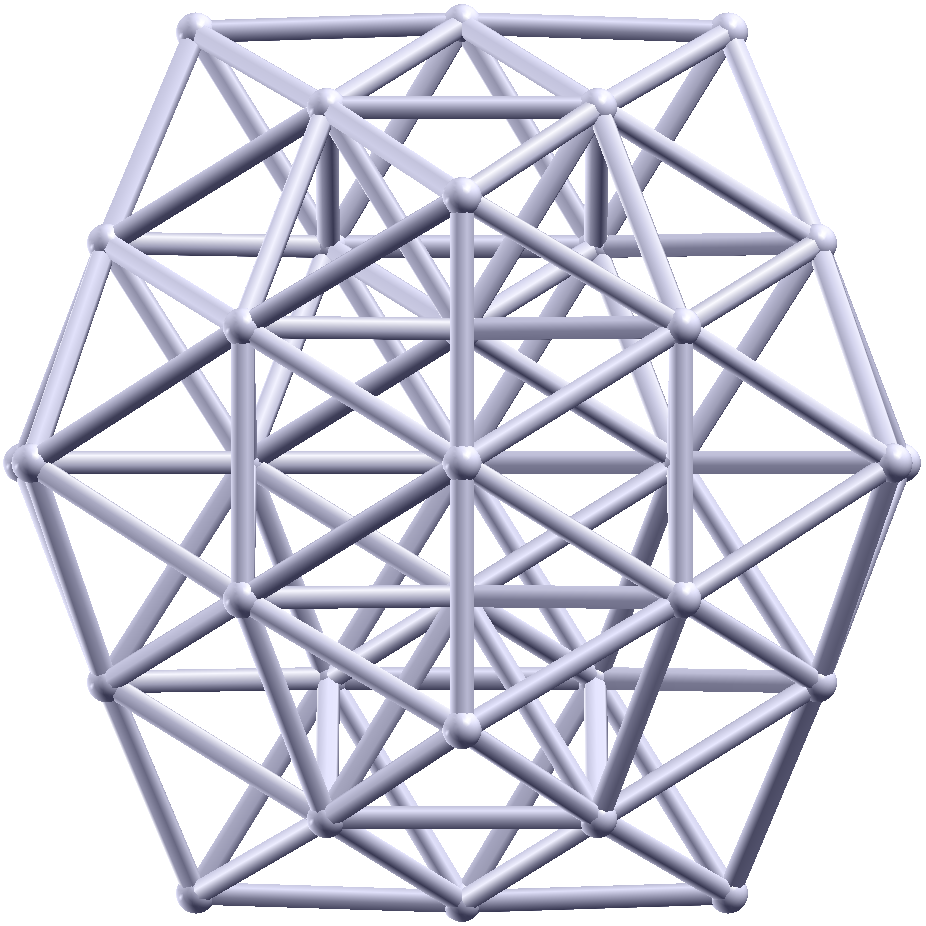} &
\includegraphics[width=4.5cm]{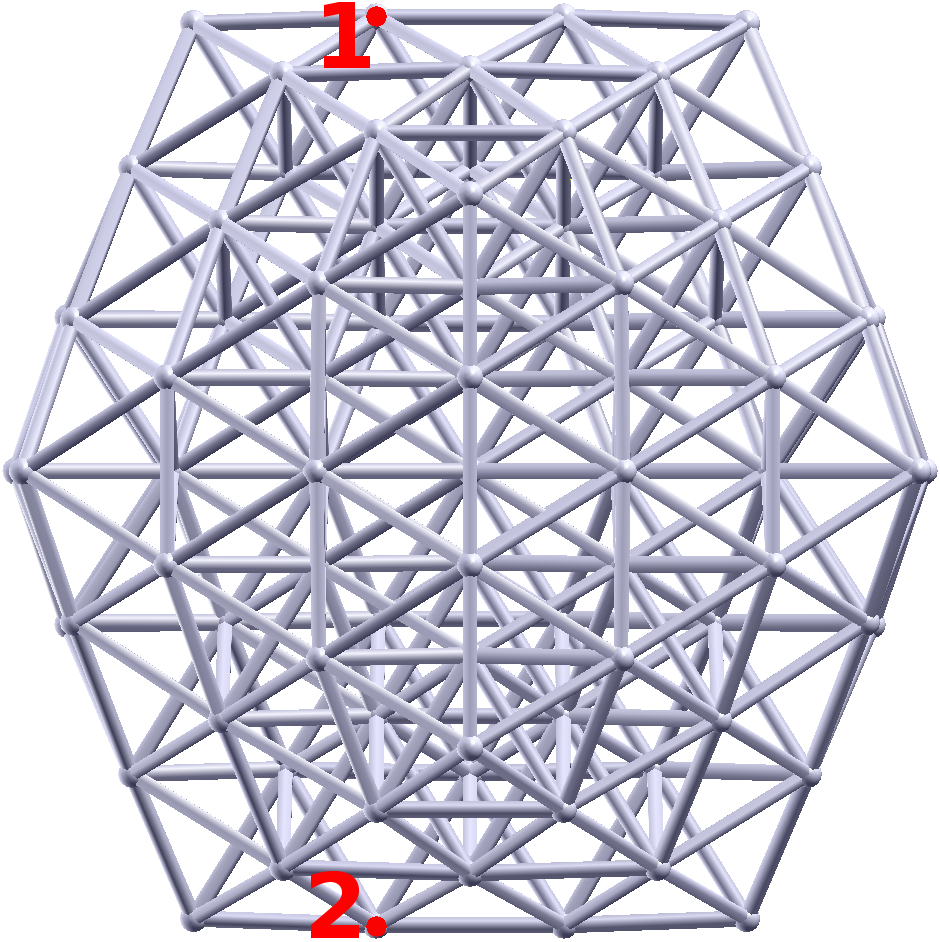} &
\includegraphics[width=4.5cm]{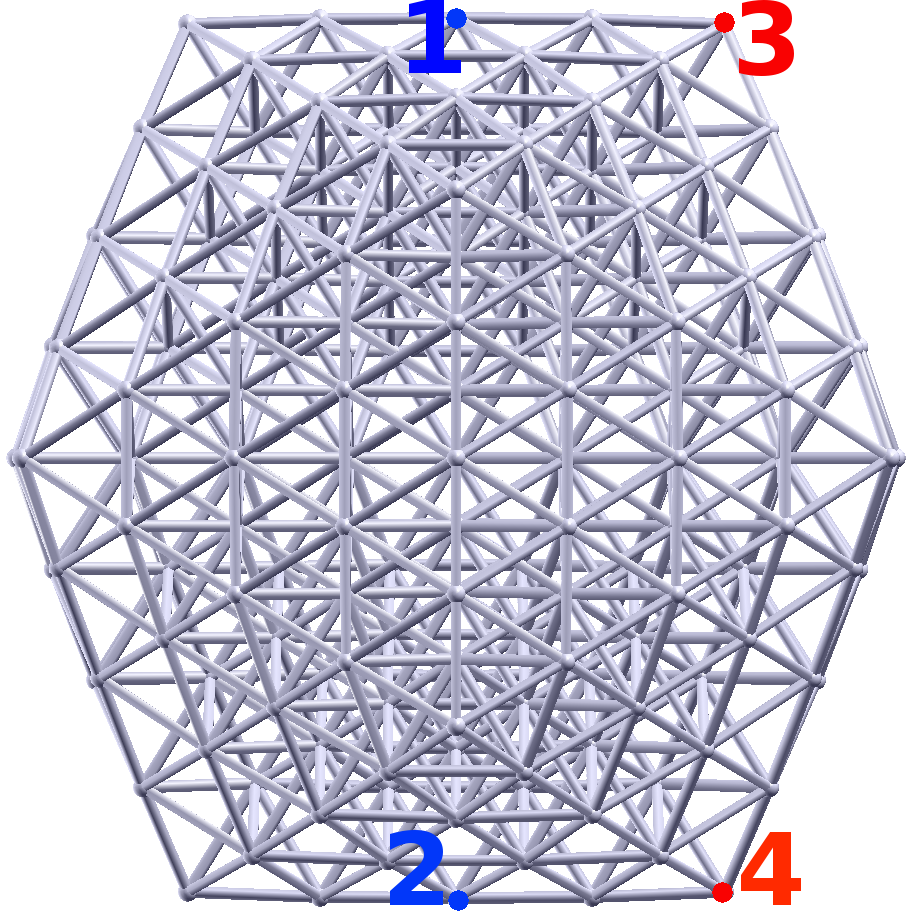} \\
\end{tabular}
\caption{\label{f:geom} Relaxed geometries of icosahedral silver clusters.
Labels S3, S4, S5 refer to the number of atom layers present 
in the clusters. Labels L1, L2 refer to the number of layers 
that are kept in a given cluster shell. The structures in the lowest 
row represent compact clusters. Atoms whose relative distances are 
discussed in text are marked by colored numbers.}
\end{figure}

A set of representative clusters and cluster shells are shown in 
figure \ref{f:geom}. We characterize icosahedral clusters by the 
number of atom layers present in the cluster and refer to this 
number as size of the cluster. Cluster shells are constructed 
starting from a cluster and keeping only several outer atomic 
layers---similar to the approach adopted by other groups 
\cite{Weissker:2014,Stener-shells:2014}.
The number of atoms in a particular layer $l>1$ is given 
by $N(l) = 10 l^2 -20 l + 12$. The first ``layer'' is composed of one atom.
For example, the largest cluster we considered is composed of six atom layers
(denoted S6L6) which makes in total 1+12+42+92+162+252=561 atoms,
while the icosahedral shell S7L4 will contain 92+162+252+362=868 atoms.

As already mentioned, we optimize the geometries of the clusters and shells
in order to account for rearrangement effects caused by surface stresses.
Geometrical relaxations were done by minimizing the total forces
as implemented in SIESTA. The DFT relaxation only slightly compressed the
ideal Mackay structures. For instance, the distance between extreme 
atoms 1 and 2
marked in figure \ref{f:geom} for the S4L4 cluster is 14.76 and 14.52 \AA{}
for the ideal geometry (in which the experimental lattice constant 4.09 \AA{} 
is used) and the Wu and Cohen geometry, respectively.

The surface stress also tends to distort (round up) ideal geometries.
For instance, the distance between atoms 1 and 2 in the S5L5 cluster 
is slightly different than the distance between extreme atoms 3 and 4
(see figure \ref{f:geom}). The distance $d_{12}$ is 19.68 and 19.40 \AA{}
for the ideal and the GGA geometries respectively, while the distance
$d_{34}$ is 19.23 \AA{} for the GGA geometry.

Both geometry distortions (overall compression and rounding) are present
in the silver shells. For instance, the GGA-relaxed length $d_{12}$ 
in the S4L1 shell is 13.79 \AA{}. Distances $d_{12}$ and $d_{34}$ in 
S5L1 shell are 18.18 and 18.14 \AA{}, respectively. Notice that these
distances are smaller than those in the corresponding compact structures,
indicating a larger compression in the case of mono-layered shells.

Summarizing the outcome of GGA geometric relaxations, we note that relaxation
leads to minor geometrical distortions of the ideal icosahedral clusters. However,
it leads to relatively large compressions for thin silver shells. For instance,
if we characterize the compression with an averaged bond length in the edge of
clusters (chosen as a simple measure and representative for the other nearest
neighbor distances in the cluster),
then the mono-layered shells have that bond length compressed
by 6\% (2.80 \AA{}) as compared to 
the average bond length in the compact clusters (2.97 \AA{}).
The cluster averaged bond length is also compressed by 
2.4\% as compared to the ideal bond length (3.04 \AA{}) calculated 
with the experimental lattice constant of bulk silver. 

%
%
%
\subsection{Optical absorption of Ag$_N$ icosahedral clusters}

Qualitatively, one expects that the plasmon resonance is affected
by the size and morphology of the clusters. In order to assess the magnitude
of this dependence, we performed TDDFT calculations of the absorption 
spectra for compact clusters and shells.
The smallest cluster is composed by two layers (Ag$_{13}$), while 
the largest cluster consists of six layers (Ag$_{561}$).
In order to assess the effect of removing internal atoms, we also performed
calculations of hollow structures, keeping up to four outer atomic
layers in a given system.

In order to expand the response and induced density
we have chosen an atom-centered
product basis as described in section \ref{s:methods} with 77 functions
per atom. The dominant product basis \cite{Foerster:2008,df-pk:2009,iter_method}
is used as an intermediate basis in the application of the 
non-interacting response, as explained above.
Depending on the system, the size of the dominant product basis 
is 2.4--4.1 times larger
than that of the atom-centered product basis. The generation of the 
product basis 
$A^{\mu}(\bm{r})$ and the calculation of interaction matrices $K^{\mu\nu}$
takes a relatively small amount of time. For instance, in the case of 
our largest Ag$_{561}$ cluster, a 12 core Intel machine spends (approximately)
45, 1 and 260 minutes, respectively for the basis set generation, and the 
calculations of Hartree- and GGA- kernels. The iterative procedure
normally takes most of the walltime. We decided to compute the absorption 
spectra in a range 0--10~eV, with a frequency step $\Delta\omega=0.05$~eV,
and a broadening constant $\varepsilon=0.08$~eV (i.e. full-width at half
maximum is 0.16 eV). This consistent choice of frequency step and 
broadening ensures that we do not ``overlook'' any feature 
present in the computed data and also produces data that can be
well interpolated.

\begin{figure}[htbp]
\begin{tabular}{p{7.5cm}p{3.5cm}p{3.5cm}}
\includegraphics[width=7.5cm]{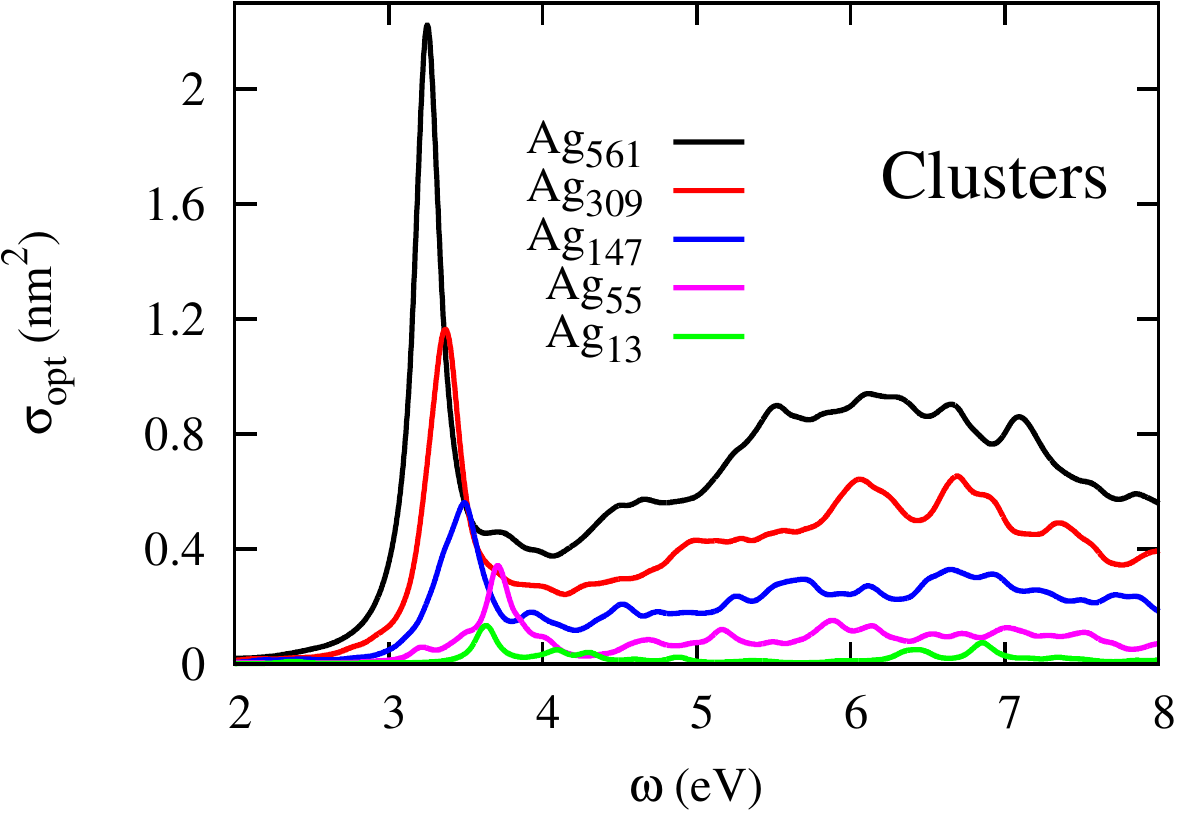} &
\includegraphics[width=3.6cm]{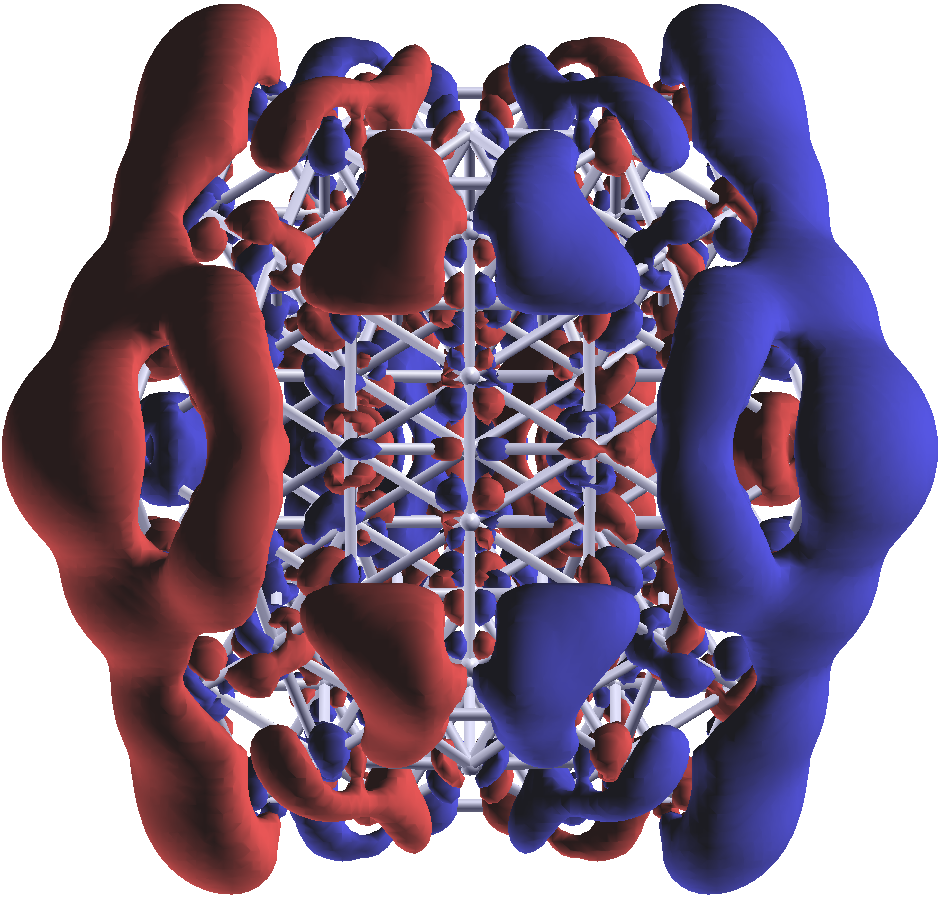} &
\includegraphics[width=3.6cm]{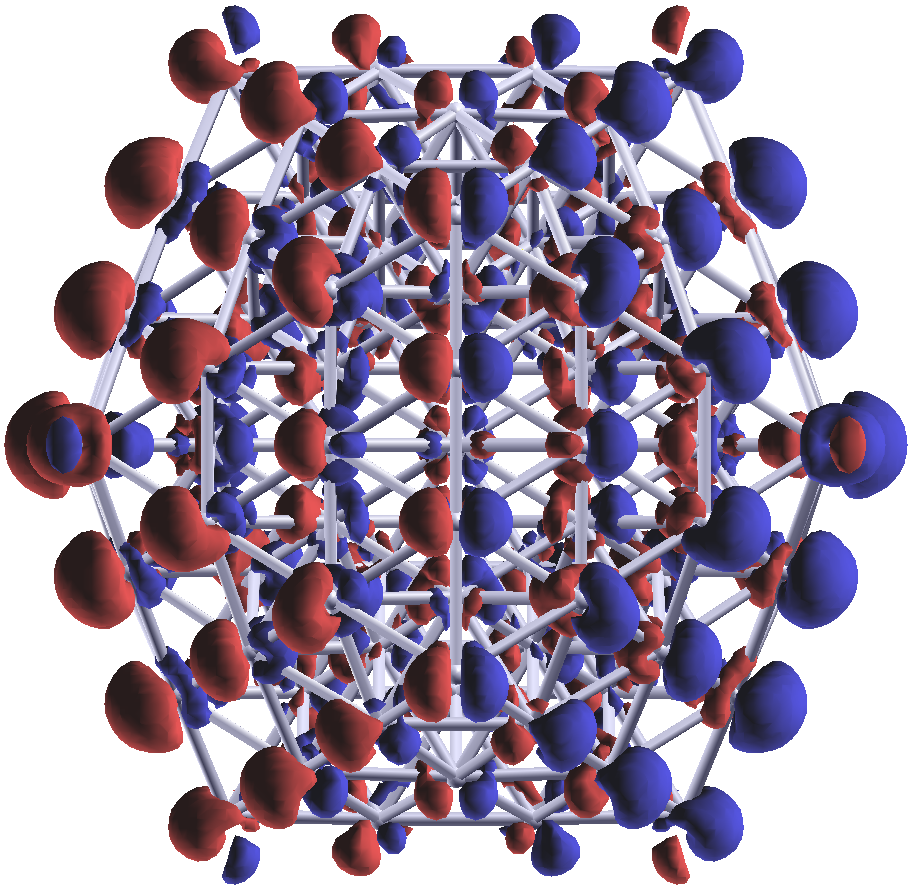} \\
a) & b) & c) \\
\end{tabular}
\caption{\label{f:abs_cls}The absorption cross sections of 
silver clusters of icosahedral shape are shown on panel a).
One can recognize resonances around 3--4 eV (sharp) and 
6--7 eV (very broad).
Panel b) and c) show the isosurfaces of density change
$\mathrm{Re}(\delta n(\bm{r},\omega))$ for the Ag$_{147}$ cluster 
at the frequencies 3.4 and 6.6 eV, respectively.}
\end{figure}

We focus our attention on the optical 
absorption properties of compact icosahedral geometries
and then will move to a comparison with silver shells
in the next subsection \ref{ss:shells-results}. 
Photo-absorption cross sections are shown in the figure \ref{f:abs_cls} (a). 
The cross section possesses two maxima: a sharp peak
at 3--4 eV and a broad maximum around 6--7 eV.
The frequencies of the resonances decrease for larger clusters. 
For instance, the two largest clusters Ag$_{309}$ and Ag$_{561}$ 
have the absorption maxima at 3.37 and 3.25 eV in the low-frequency band 
and at 6.7 and 6.2 eV in the high-frequency band, respectively.
Panels b) and c) in 
figure \ref{f:abs_cls} show the induced density change 
(solution of equation (\ref{dn-pb})) in the Ag$_{147}$ cluster
at the maxima of the low-frequency and high-frequency bands,
respectively. The direction of external field is set along x axis,
i.e. collinear with the plot plane and horizontal. The isosurfaces of the
real part of $\delta n(\bm{r},\omega)$ were plotted for a 10\% of
the corresponding maximal value. One can see that the 
low-frequency resonance is caused by an oscillation of charge with 
pronounced dipole character, while the high-frequency band is supported
by more localized, homogeneously-spread oscillations. 
The lower energy resonance obviously corresponds to the 
dipole Mie plasmon of the particle.
From a quantum mechanical
point of view, both bands consist of many nearly-degenerate transitions.
The number of the transitions makes it impractical to analyse each
of them in detail. However, one can provide some analysis
of the electronic transitions and atom-layer contributions
as demonstrated below in subsection  
\ref{ss:analysis-results}.

\subsection{Optical absorption of silver shells\label{ss:shells-results}}

\begin{figure}[htbp]
\begin{tabular}{p{7.5cm}p{7.5cm}}
\includegraphics[width=7.5cm]{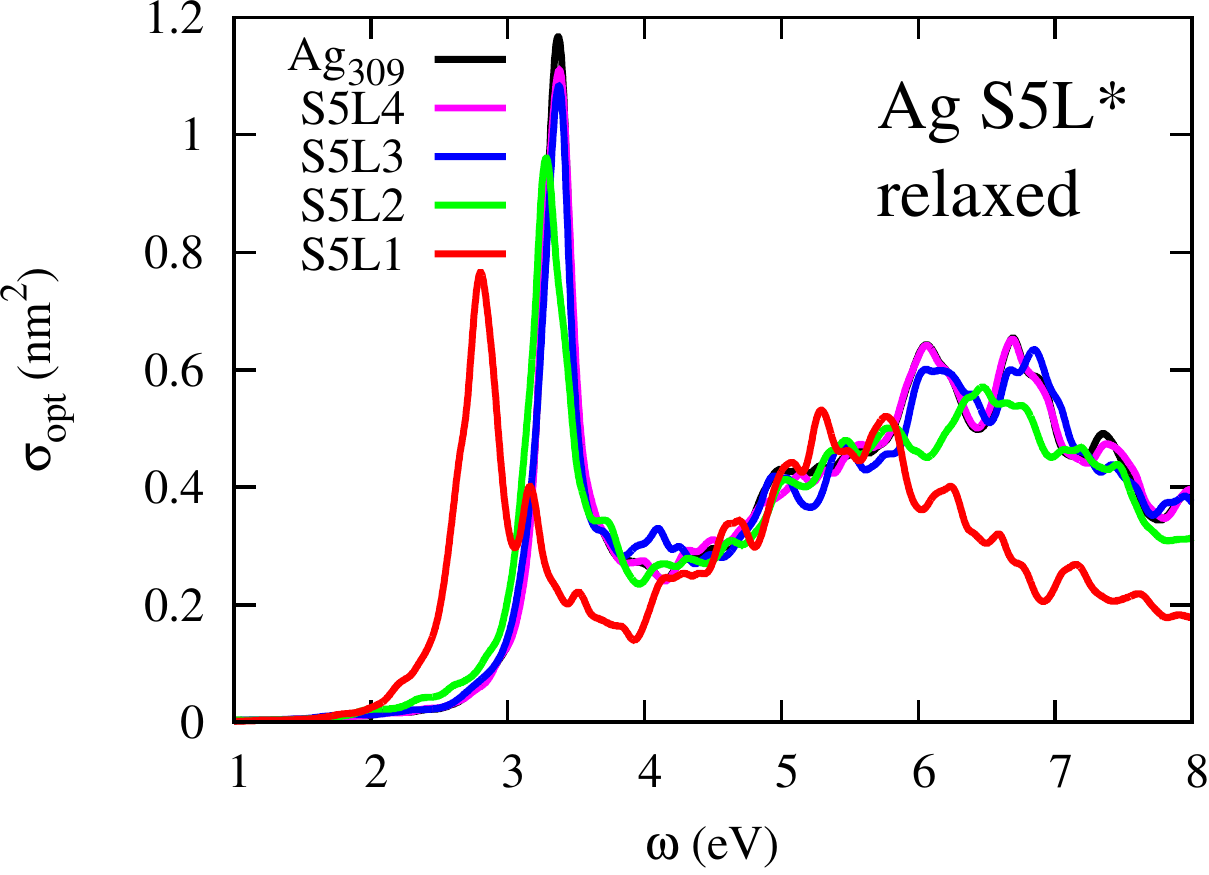} &
\includegraphics[width=7.5cm]{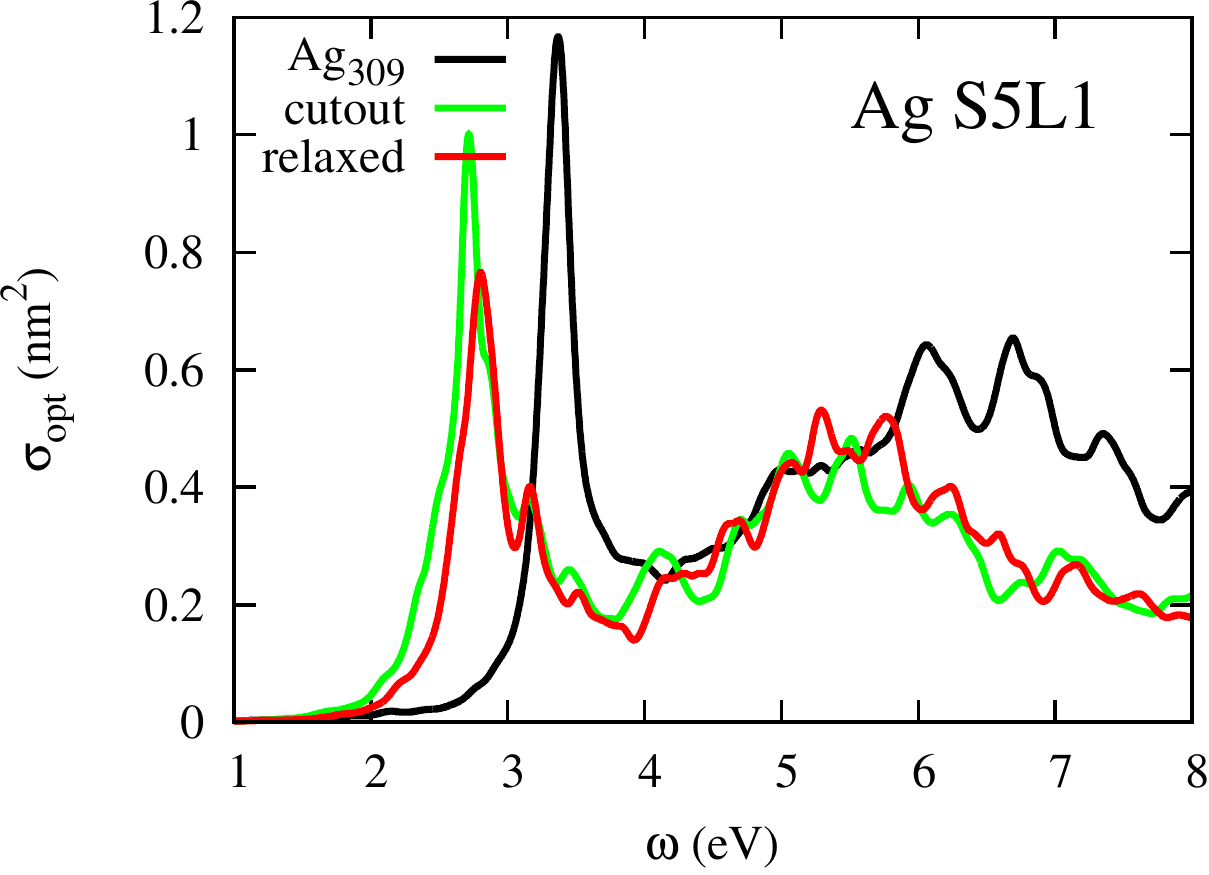} \\
a) & b) \\
\includegraphics[width=7.5cm]{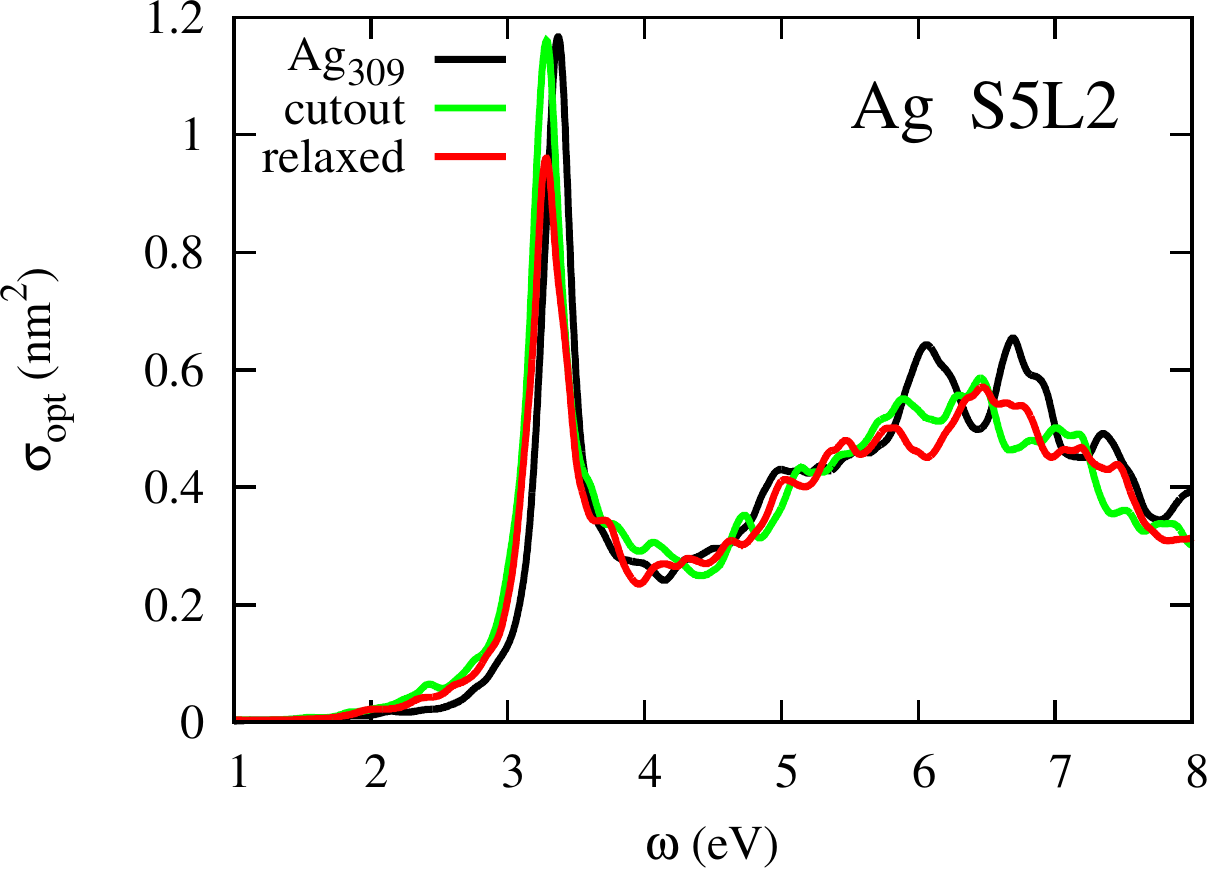} &
\includegraphics[width=7.5cm]{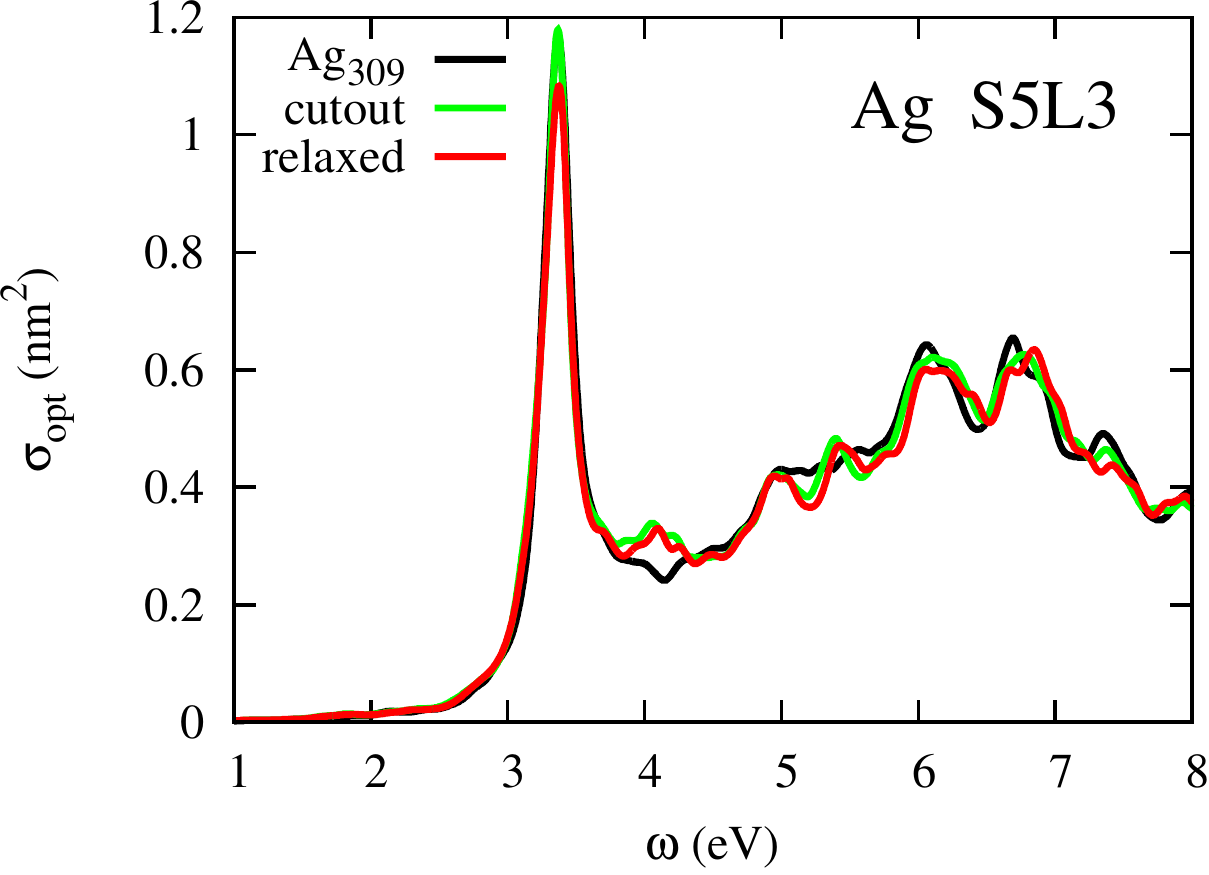} \\
c) & d) \\
\end{tabular}
\caption{\label{f:shells5}The absorption cross sections of 
icosahedral silver shells constructed from the Ag$_{309}$ cluster.
On panel a) the cross section of the Ag$_{309}$ cluster is shown
together with the cross sections of the one- to four-layered shells.
On panel b) the optical absorption cross sections of ``cutout'' 
and relaxed single-layered silver shells are compared (see text for details).
Panels c) and d) show the cross sections of 
cutout and relaxed two- and three-layered silver shells, 
respectively.}
\end{figure}

Silver shells can be constructed from the initial geometry of 
the corresponding cluster in several ways. We suggest two useful ways to
construct the silver shells. In the first way, we simply delete atoms
of several inner atomic layers from the compact cluster
that was previously relaxed within DFT. In the second way, we additionally 
relax the positions of the atoms remaining in the shell. The former way is
less computationally demanding and also is eventually more useful
to approximate the response of the whole cluster by the response of its
outer shell. The latter way should give results that are closer to the
corresponding experimental values for shells. In figure \ref{f:shells5}, we
analyse the optical absorption cross section of the Ag$_{309}$ cluster and
of silver shells derived from that cluster. In panel a) we show 
the cross sections of the relaxed shells. It is worth noting that
a single-layered shell (S5L1, Ag$_{162}$) exhibits a low-frequency
resonance at 2.8 eV which must be compared to the 3.37 eV in the case of 
the Ag$_{309}$ cluster. If we leave two atomic layers as in the silver shell
S5L2 (Ag$_{254}$), then the low-frequency band shifts back (3.28 eV) almost
to the frequency of the full cluster. The broad high-frequency 
resonance of the silver shells
qualitatively follows the same behavior. Namely, the maximum frequency of the
single-layered shell is red shifted to 5.8 eV that must be compared to 6.7 eV
in the case of Ag$_{309}$ cluster. However, already a two-layered shell almost
recovers (6.5 eV) the position of the maximum of the high-frequency band
for the compact cluster. The optical cross sections 
of the three- and four-layered shells approach that of the cluster steadily.

\begin{figure}[htbp]
\begin{tabular}{p{7.5cm}p{7.5cm}}
\includegraphics[width=7.5cm]{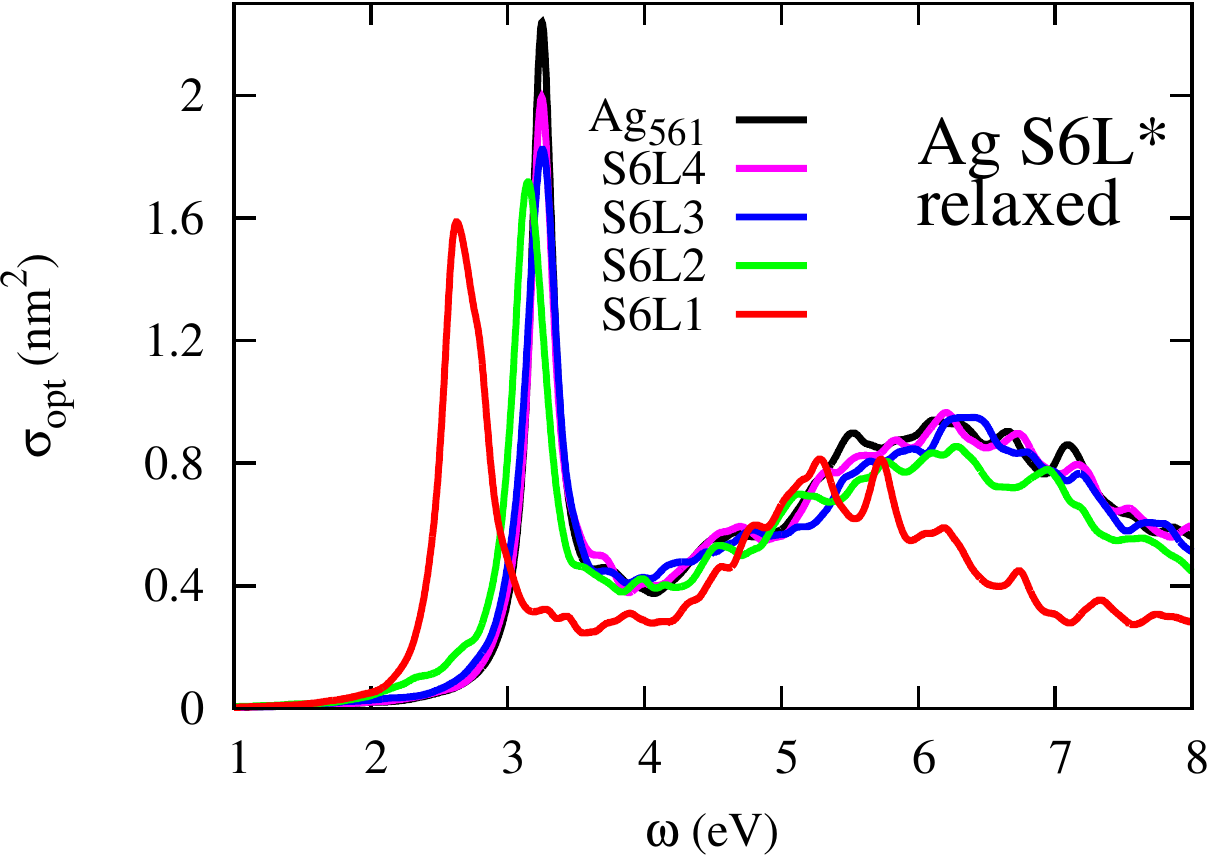} &
\includegraphics[width=7.5cm]{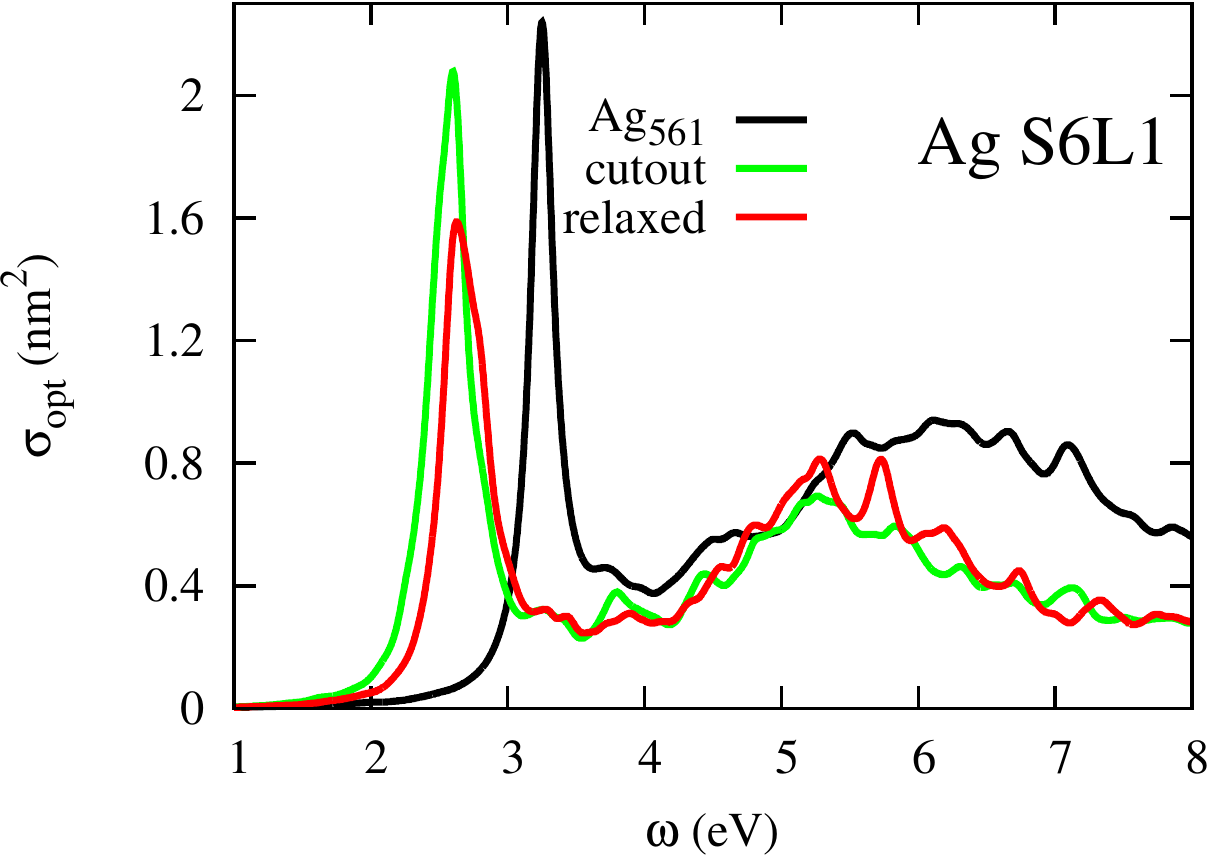} \\
a) & b) \\
\includegraphics[width=7.5cm]{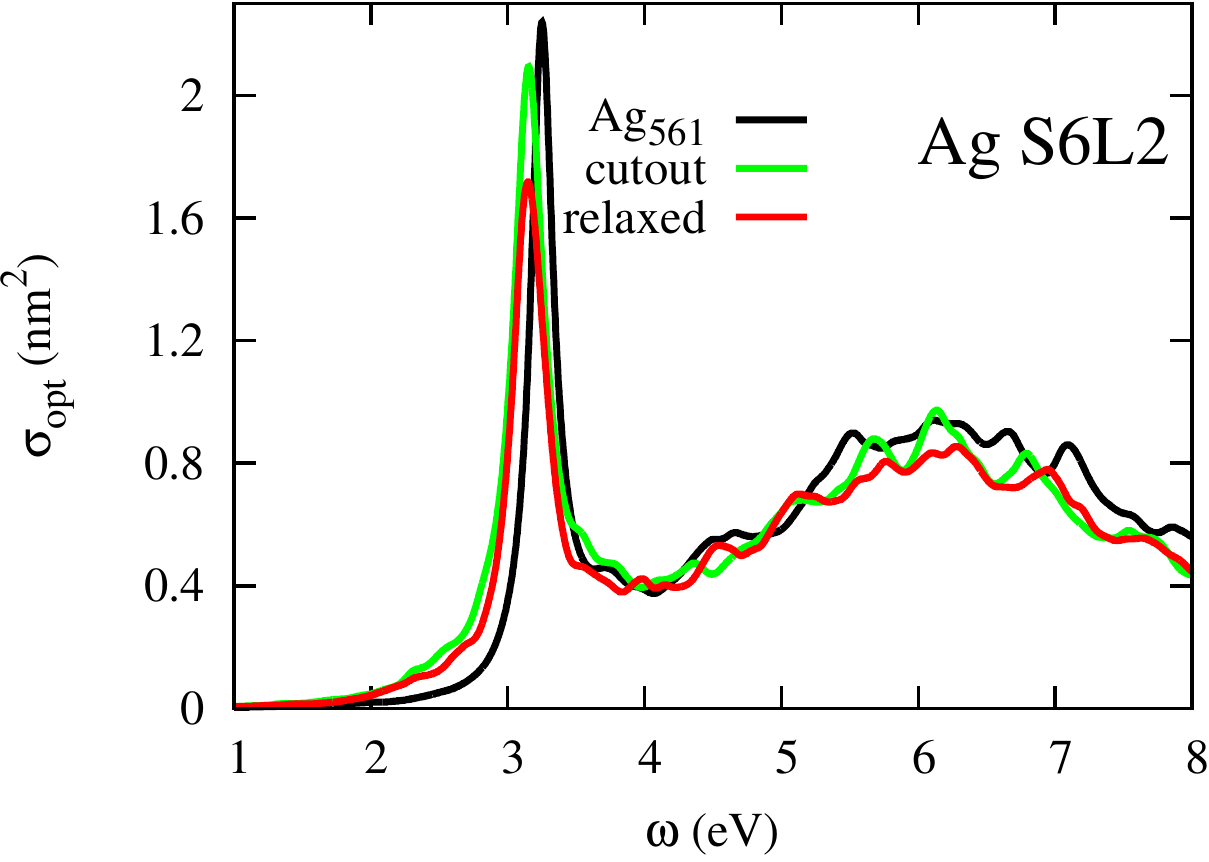} &
\includegraphics[width=7.5cm]{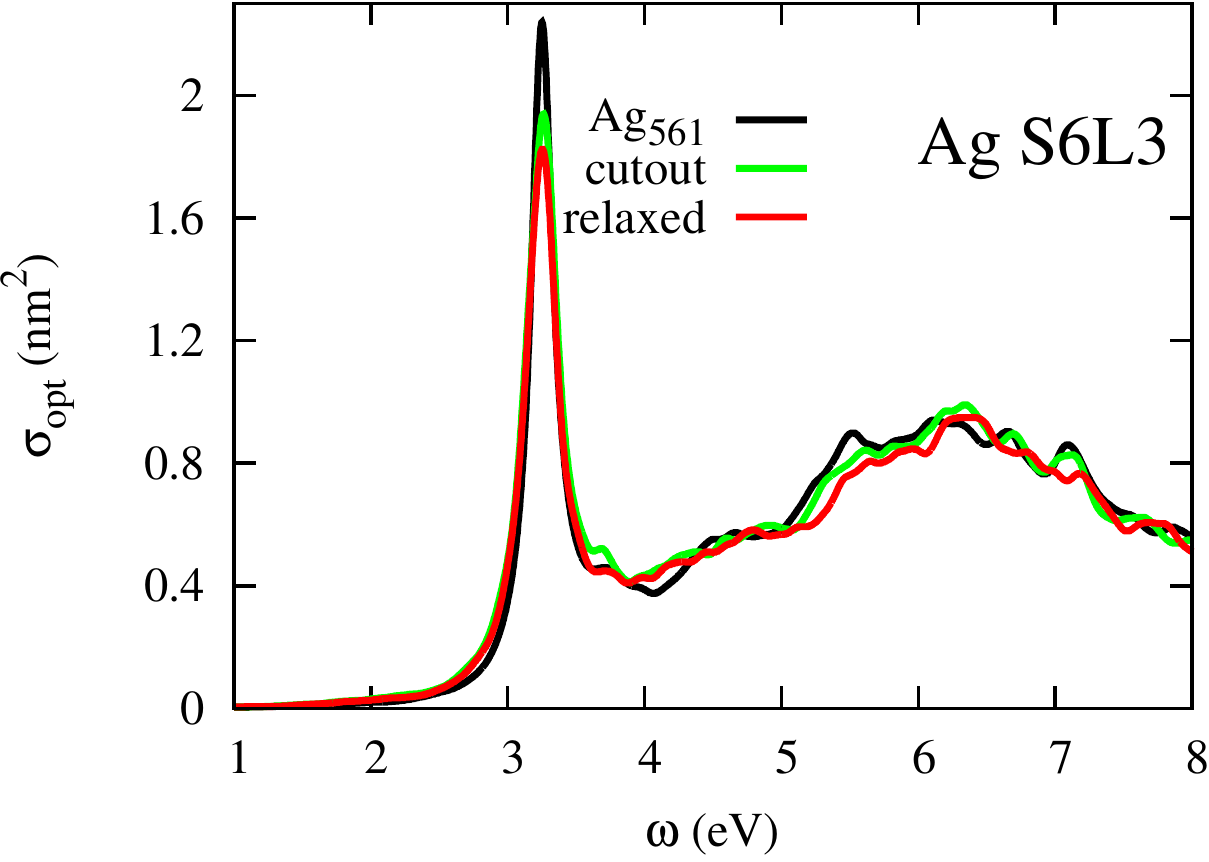} \\
c) & d) \\
\end{tabular}
\caption{\label{f:shells6}The absorption cross sections of 
icosahedral silver shells constructed from the Ag$_{561}$ cluster.
On panel a) the cross section of the Ag$_{561}$ cluster is shown
together with the cross sections of the one- to three-layered shells.
On panel b) the optical absorption cross sections of cutout and
relaxed single-layered silver shells are compared. Panels c) and d) show
the cross sections of cutout and relaxed two- and three-layered silver
shells, respectively.}
\end{figure} 

The effect of geometry relaxation in the silver shells is also quantified
in figure \ref{f:shells5}. The panels b), c) and d) show the
optical absorption cross sections of the one-, two- and
three-layered silver shells, the geometry of which was either ``cutout''
from the geometry of the Ag$_{309}$ cluster or optimized (relaxed). 
The comparison shows that the effect of 
geometry relaxations is the much larger for the single-layered shell
than for the other shells. The 
relaxation leads to slight blue shift of both bands.
This is due to the additional compression of the structure when
relaxed.
Moreover, relaxations result in a broadening of the
low-frequency resonance and in an relative increase of the cross section
in the high-frequency band in the single-layered shell (panel a).
Panels c) and d) show that the effect of geometry relaxation is
less important for thicker shells. 
Cutout structures, created by simply removing the internal layers
of atoms, give results very similar to those of the relaxed shells.

\begin{figure}[htbp]
\begin{tabular}{p{7.5cm}p{7.5cm}}
\includegraphics[width=7.5cm]{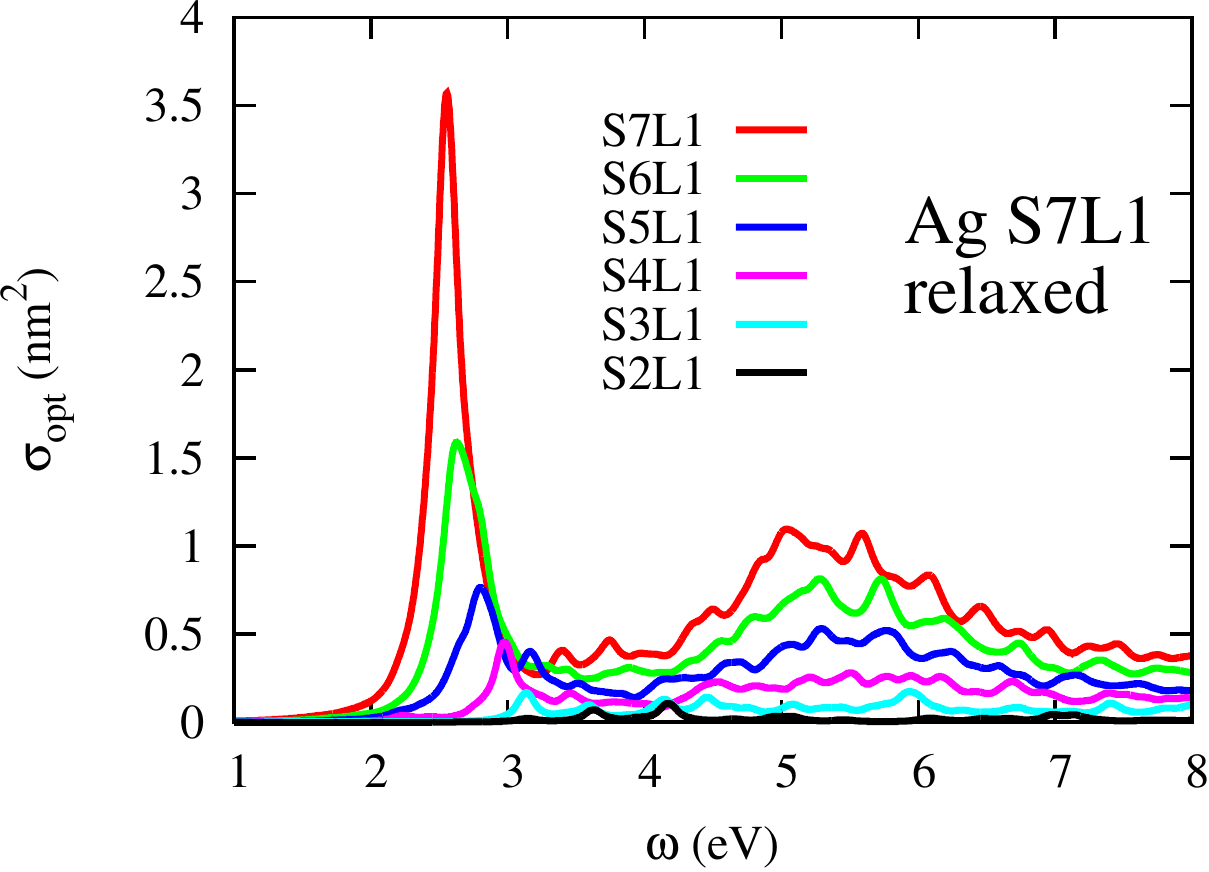} &
\includegraphics[width=7.5cm]{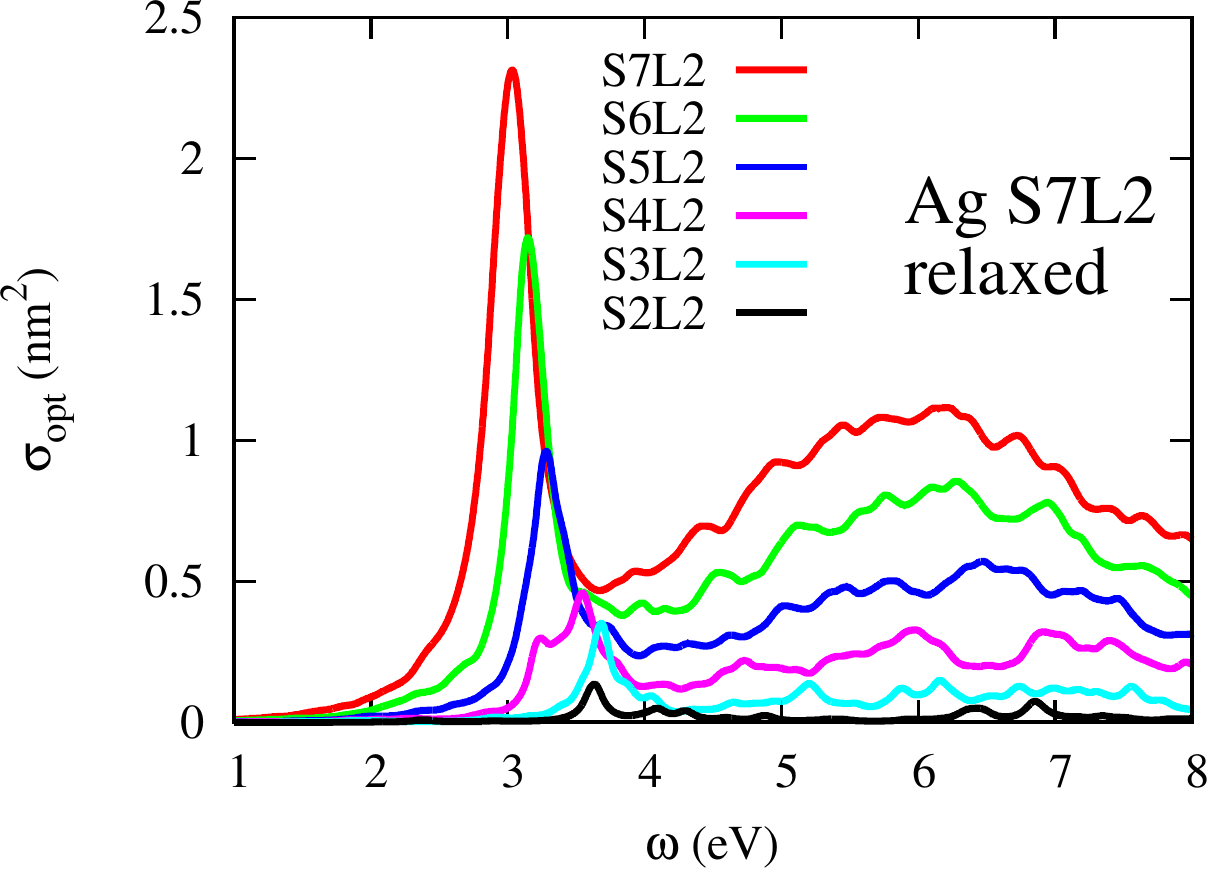} \\
a) & b) \\
\includegraphics[width=7.5cm]{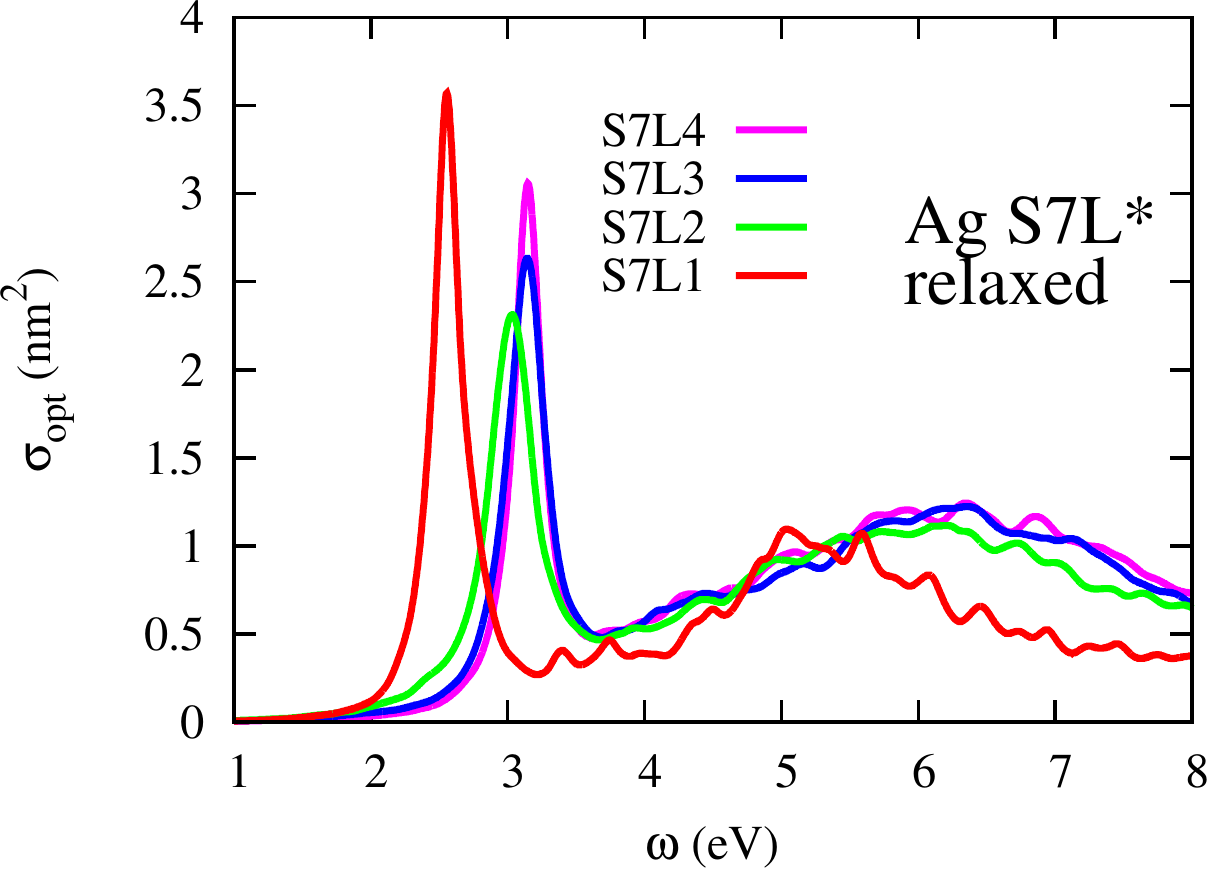} &
\includegraphics[width=7.5cm]{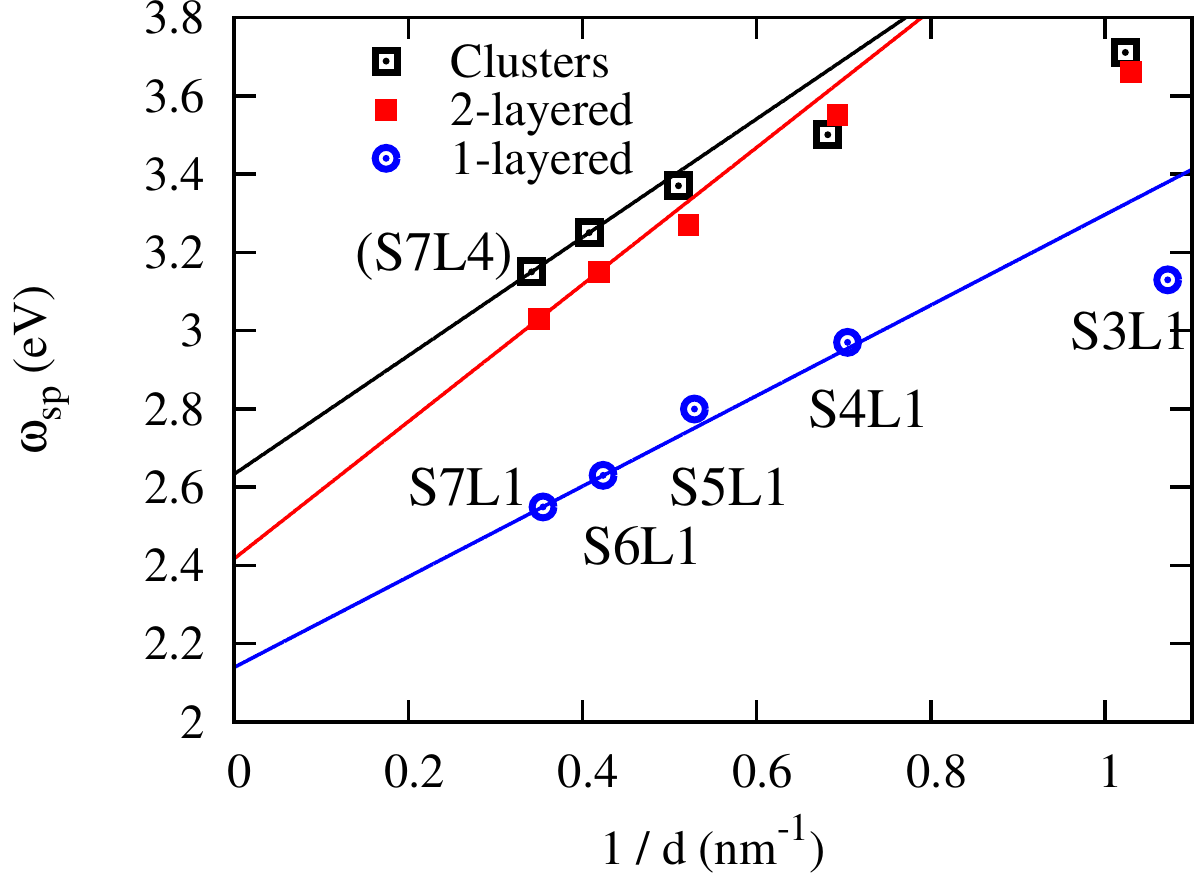}
\\
c) & d) \\
\end{tabular}
\caption{\label{f:shells7}The absorption cross sections of the 
single- (panel a) and double-layered (panel b) icosahedral silver shells.
On panel c) the absorption cross sections of shells of size 7 are compared.
The plasmon frequency versus size of the clusters and shells on panel d).
Lines on panel d) are drawn through the points corresponding to largest 
structures (S6 and S7).}
\end{figure} 

In figure \ref{f:shells6} we show the optical absorption cross
sections of the 6-layered icosahedral silver cluster and shells constructed
from that cluster. The cross sections show qualitatively the same behavior
as those of the 5-layered cluster and shells 
(see figure \ref{f:shells6} panels a), b), c) and d)). Namely, 
the cutout and relaxed one-layered shells S6L1 have the red-shifted 
low-frequency resonance at 2.6 and 2.62 eV, correspondingly, while the 
cross sections  of two- and three-layered shells are much closer 
to that of the full cluster (3.25 eV). However, the low-frequency band 
of the two-layered shells S6L2 differs more from that of 
the compact cluster (3.15 eV for cutout and relaxed geometries).
It is also interesting to note that geometry relaxations of the one- and 
two-layered shells lead to a small blue shift relatively to the 
unrelaxed calculations.
In the case of three-layered shell S6L3, however,
the geometry relaxations lead to 
a slight red shift (3.25 eV) of the low-frequency maximum in comparison
to the ``cutout'' geometry (3.26 eV).

In figure \ref{f:shells7} we collected the optical absorption of 
all single- (panel a) and double-layered (panel b) shells computed in
this work. The cross section of the shells is qualitatively similar to
the cross section of the clusters: there are low-frequency sharp
and high-frequency broad resonances, the maxima of these resonances
steadily decreases with increasing the cluster size. 
The ratio of low-frequency to high-frequency intensities is nearly
the same in compact structures and shells of sizes up to 6,
while for the largest single-layered shell S7L1 the relative
intensity of the low-frequency peak increases. 
The low-frequency peak in the absorption cross section of the S7L1 shell 
is more intense than for the S7L2 shell (figure \ref{f:shells7}, panel c),
in contrast to what is observed in smaller shells
(figures \ref{f:shells5} and \ref{f:shells6}). The absorption cross
section of the thicker largest shells (S7L3 and S7L4) should approach the 
response of the compact cluster of the same size. Unfortunately, we could
not compute the absorption of the cluster Ag$_{943}$, because of the 
large memory requirements, but we can be reasonably sure that 
the thickest shell S7L4 represents well the absorption of the compact
cluster of size 7. The frequency of maximal absorption shows
an approximately linear dependency on the inverse diameter of clusters 
as shown on
panel d) both for clusters and shells. 
The effective diameter of the clusters and shells has been computed
from the spatial cross-section of the cluster, while the diameter of the 
cluster is defined as the diameter of a circle with the same area. 
The data shown in figure  \ref{f:shells7} (d) indicates
small discrepancies from the linear trend both for compact geometries
and double-layered shells for smaller clusters, while in the case of
single-layered shells such discrepancy is visible only for smallest shell.

\subsection{Analysis of the optical absorption\label{ss:analysis-results}}

\begin{figure}[htbp]
\includegraphics[width=14cm]{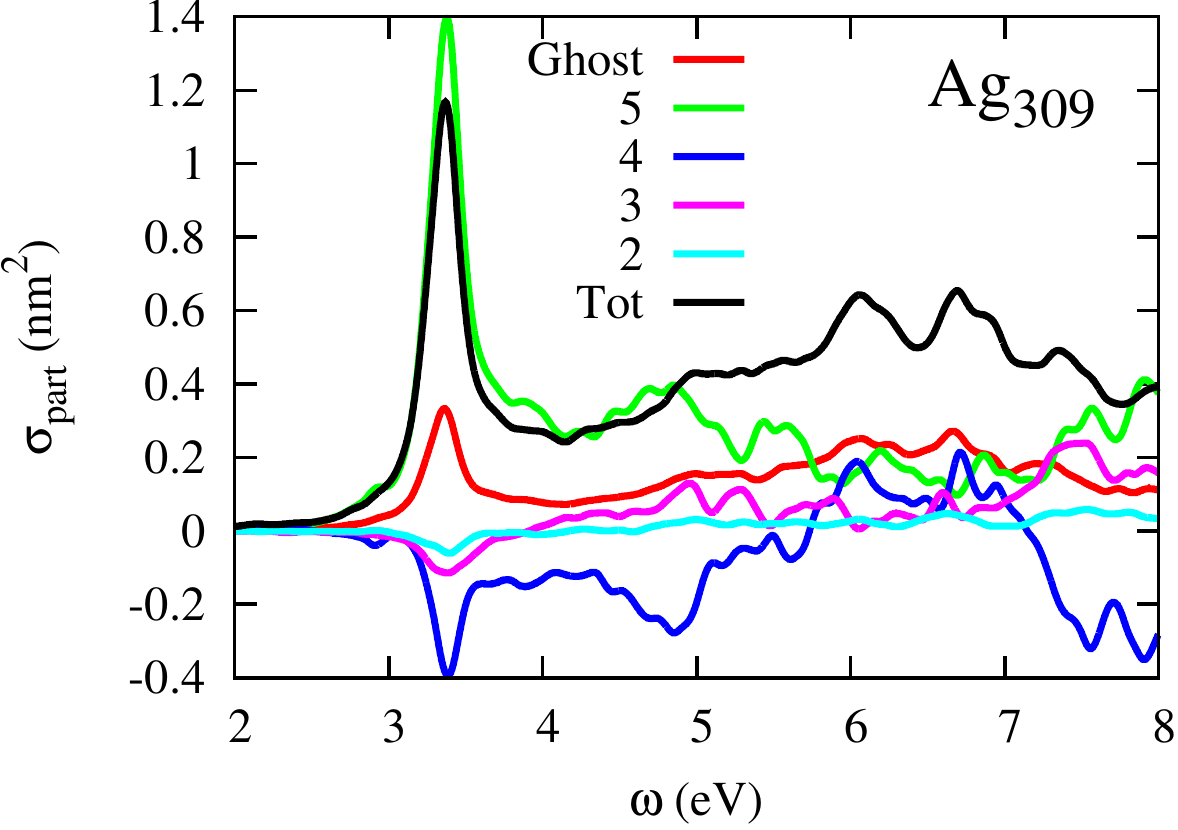} 
\caption{\label{f:tags}
Contribution of the different atom layers 
to the absorption cross section of the 
Ag$_{309}$ cluster (S5L5).}
\end{figure}

In figure \ref{f:tags} we show the analysis of the 
atom-layer contributions to the optical absorption cross section
according to the method presented above in section \ref{ss:analysis}.
The partial cross sections corresponding to different atom-layer 
polarizabilities (\ref{pol_tags}) are plotted
together with the total absorption cross section. The outer layer 
(5-th layer, 162 atoms) gives rise to the largest partial cross section
that is even larger than the total cross section at resonance.
The 6-th layer of ``ghost atoms'' is contributing constructively
in the whole frequency range and with a rather large magnitude 
that is approximately equal to that of the second inner layer
(4-th layer) of ``true'' atoms. The inner atomic layers give rise
to negative partial cross sections at least in some frequency ranges.
At the main resonance 3.37 eV, all inner layers contribute
destructively. The second outer layer (4-th layer, 92 atoms) gives rise 
to a negative partial cross section in most of the frequency range
considered. The contribution of the first ``layer'' that is
composed of one atom is much lower than that of the next layer
(2-nd layer, 12 atoms) and is not shown. 

The interpretation of these results in the low-frequency 
range is quite clear. As expected from the plot in figure \ref{f:abs_cls} (b),
and consistent with the Mie plasmon character of the resonance,
the main contribution is coming from the surface layer.
For this mode, it is critical to describe accurately the polarization
of the surface, which explains the large contribution of the layer
of ``ghost atoms'' above cluster surface. Those basis orbitals 
control the extension of the electronic states towards vacuum
and, therefore, are instrumental to correctly account for the polarizability
of the surface. The change of signs is related to the relative phases of 
the different contributions, reflecting to some extent the global
nodal structure of the contributing electronic states.

At higher energies, the smaller dielectric function of bulk silver makes
the charge screening less efficient, the modes loose their 
predominant surface character, and the contributions
of different layers become more similar in intensity.

\begin{figure}[htbp]
\includegraphics[width=14cm]{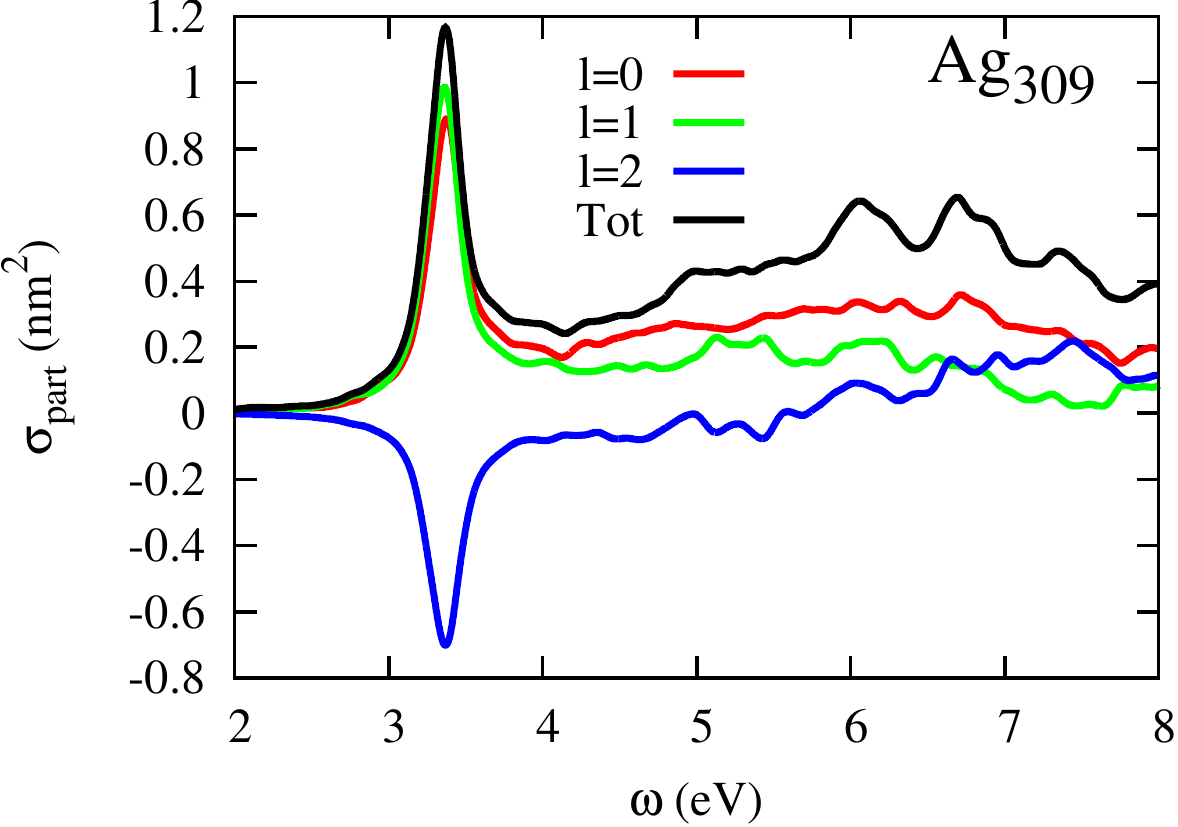}
\caption{\label{f:occ_virt}Analysis of the different angular momenta
of the basis orbitals involved in the description of the occupied states 
contributing to the absorption cross section of the
Ag$_{309}$ (S5L5) cluster as function of frequency.}
\end{figure}

In figure \ref{f:occ_virt} the partial cross sections (corresponding to the 
partial polarizabilities (\ref{pp-occ-virt})) for atomic orbitals
of different symmetry expanding the \textit{occupied} states
are shown for Ag$_{309}$ cluster.
We see that, on the absolute scale, the contribution of each angular momentum channel
is important even at low frequencies. However, there is a striking 
difference between $s$, $p$ and $d$
contributions to the absorption cross section. 
Namely, the $s$- and $p$- channels contribute
constructively to the absorption cross section, while $d$ channel contributes
destructively in the low frequency range (3--4 eV) and only starts 
to contribute constructively at high frequencies, starting approximately at 5.5 eV.
This conclusion is similar to that presented 
in a previous LDA study for smaller silver clusters
\cite{yabana:1999}. This theoretical analysis supports the view of the low-frequency 
excitation as produced by $5s$ electrons \cite{otto-petri:1976} only partially.
This is because the occupied states are $sp$-hybridized close to Fermi energy
and also there is a non-negligible contribution of $s$-symmetrical density at
higher frequencies (6--7 eV). From the other side, the strong over-screening 
of the $s$ plasmon due to $sp$-$d$ interband transitions
has been discussed at length in the literature 
\cite{Stener-gold:2011,gga-vs-lrc:2014}. The comparison
of our data with those in Refs.~\cite{gga-vs-lrc:2014,gpaw-prop:2015} 
suggests that GGA could underestimate the intensity 
of $s$ plasmon by a factor about $1.5$.

In order to comprehend better the outcome of \textit{ab-initio} modeling we 
compare our results with the available relevant
literature in the following section.

\section{Discussion\label{s:discussion}}

\begin{figure}[htbp]
\begin{tabular}{p{7.5cm}p{7.5cm}}
\includegraphics[width=7.5cm]{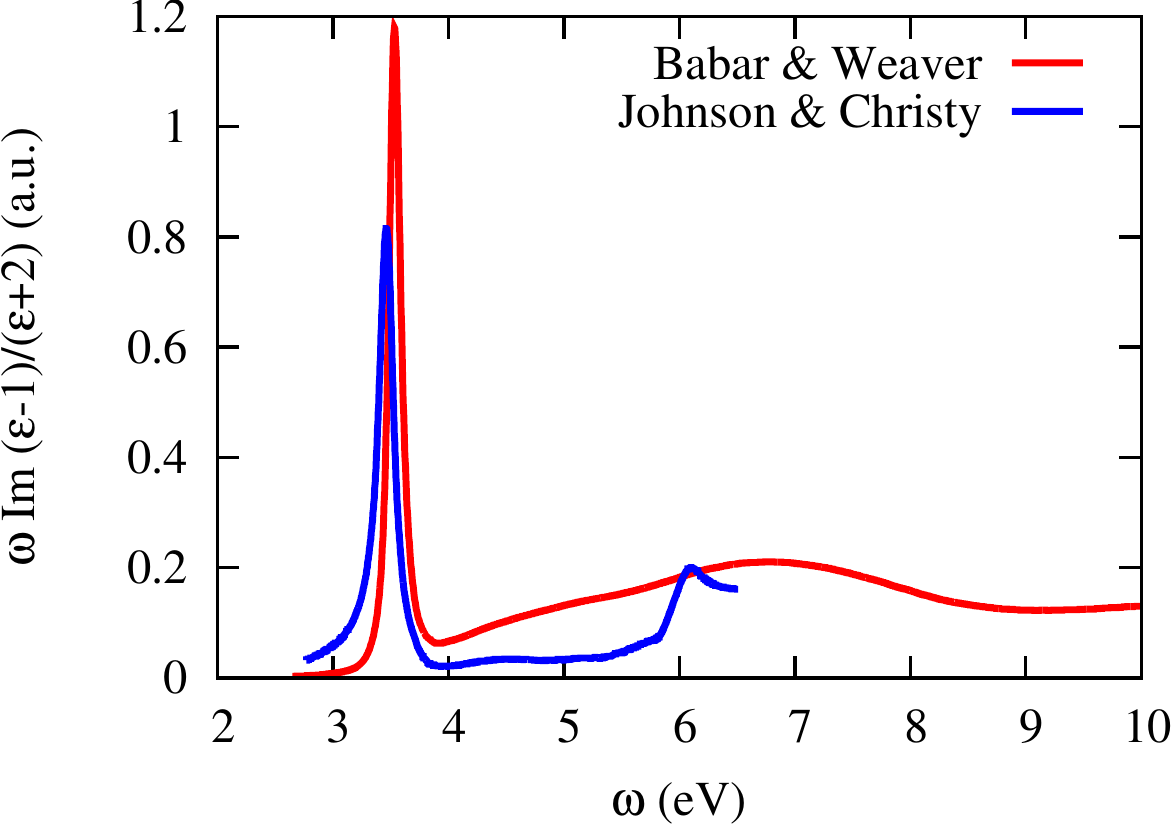} &
\includegraphics[width=7.5cm]{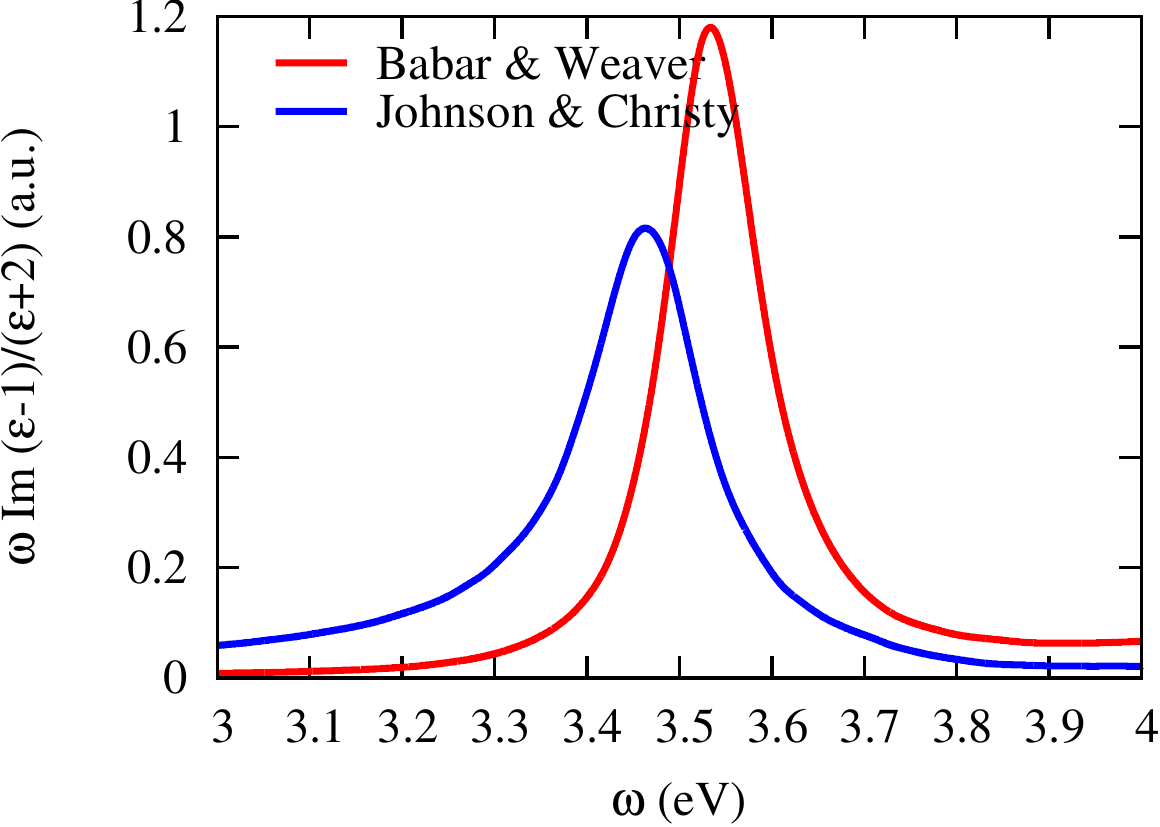} \\
a) & b) \\
\end{tabular}
\caption{\label{f:sigma_mie}The absorption cross section
in a classical quasi-static limit for a silver sphere in vacuum. 
Experimentally-determined dielectric function of silver is used.}
\end{figure}

Because the clusters that we have considered are rather large, we expect 
that their properties will approach the properties of classical 
Mie spheres \cite{Mie-book-Kreibig:1995,yabana:1999} to the extent
that the non-sphericity of icosahedra and charge spillage effects allow.
The classical absorption cross section in the electrostatic limit
depends on the dielectric function
of the material $\epsilon(\omega)$ and the dielectric function of the
embedding medium $\epsilon_m(\omega)$ \cite{Lucas:1994,Mie-book-Kreibig:1995}

\begin{equation}
\sigma(\omega) \sim \omega\, \mathrm{Im}
\frac{\epsilon(\omega)-\epsilon_m(\omega)}{\epsilon(\omega)+2\epsilon_m(\omega)}.
\label{sigma_mie}
\end{equation}
If we plot the cross section (\ref{sigma_mie}) with the experimentally determined
dielectric function of silver \cite{Johnson-Christy:1972,Babar:2015}, then 
we will get a sharp low-frequency resonance and a less intense and broad 
high-frequency resonance as it is shown in the figure \ref{f:sigma_mie} panel a).
The position of the resonances slightly varies with 
the set of experimental data used.
The more recent experiment \cite{Babar:2015} gives resonance maxima
at 3.53 and 6.8 eV, while the older and widely 
cited experimental results \cite{Johnson-Christy:1972}
give them at 3.47 and 6.1 eV, correspondingly. 
The response in the low-frequency part of 
the spectra is detailed in the panel b) of the figure \ref{f:sigma_mie}.

Comparing the quasi-static absorption cross section (\ref{sigma_mie})
with our results, we see similarities and discrepancies. First of all,
the resonance frequencies from Mie theory do not depend on the 
size of the object, but only on its shape. Classical estimations 
of the shape influence  \cite{gpaw-prop:2015} lead to a conclusion 
that icosahedral particles have to resonate at 2.9\% lower frequencies
than perfectly spherical clusters.

TDDFT accounts for charge spillage at surfaces (i.e., the fact that 
surfaces are not perfectly sharp, as assumed usually in classical 
electrodynamics) which leads to a divergence of the plasmon 
frequency $\omega_{\mathrm{sp}}$ from a linear dependence
$\omega_{\mathrm{sp}} = k d^{-1} + b$ with respect to the 
inverse diameter
for small clusters (see figure \ref{f:shells7} d).
Measurements of the optical absorption of small silver clusters 
support this conclusion at least qualitatively 
\cite{Liebsch:1993,Ruppin:1993,Charle:1998,Haberland:2013}.
The divergence from the linear dependence in the 
experimental data must be due to charge spillage, because the other
effect that leads to red shifts
-- higher multipole contributions to the electron-photon coupling --
does affect only relatively large clusters of diameters 50 nm and
larger \cite{Kreibig-multipoles:1970}.

In the case of the smallest cluster presented here Ag$_{13}$, there
are experimental data \cite{ag13-in-ar:1993,ag13-in-ar:2008}
for argon embedded clusters that show a maximum absorption strength at 
3.4-3.5 eV while we get 3.63 eV.
Because the effect of argon embedding can lead to red shifts as large
as 0.3--0.5 eV \cite{Matrix-effects:2006,Haberland:2013}, 
we estimate that TDDFT with GGA functional
might be delivering red-shifted frequencies by about 0.2 eV.
In the case of larger clusters, we can compare with experimental data 
\cite{Charle:1998} measured on free silver clusters.  For instance, for 
a cluster diameter of 1.2 nm, which corresponds approximately to 
our Ag$_{55}$ cluster, 
the experimental peak position is 3.51 eV, while we get 3.7 eV.
For a cluster diameter of 2.2 nm, which corresponds to
our Ag$_{309}$ cluster,
the experimental peak position is 3.47, while we get 3.37 eV.
For a cluster diameter of 2.8 nm, which corresponds to
our Ag$_{561}$ cluster,
the experimental peak position is 3.45, while we get 3.25 eV. 
These comparisons
indicate that our calculations deviate by no more than 0.2 eV
from experiment, presenting a somewhat stronger size dependency
than observed experimentally. The computed plasmon frequencies are 
red-shifted for large clusters.

Besides experimental results, there are many calculations of silver 
optical absorption properties available in the literature. 
The first observation is that very little can be found on the
broad resonance in the frequency range 5.5--7.5 eV,
although this feature must be available in the calculations done 
with atomistic codes. This is probable due to little practical relevance
of the frequency range and the much stronger low-frequency plasmonic response.

The position and strength of the low-frequency plasmonic peak depends on the 
used functional and the basis set: the agreement with experiment improves 
while using more sophisticated functionals and larger basis sets. For instance,
the use of the simplest LDA functional results in 
a too low energy onset of the $d$-bands in 
the electronic structure, \cite{Marini-lda-gw:2002} which leads to a reduced
strength of the low-frequency resonance and its red shift as compared to 
experiment and calculations based on Hedin's $GW$ band structures
\cite{Marini-lda-gw:2002}. In this paper, we showed that GGA
functional by Wu and Cohen 
produces a strong low-frequency peak for icosahedral clusters.
GGA functionals have been extensively used in the past for silver cluster, 
\cite{ag13-in-ar:2008,octopus-ag-au:2013} as well as the more sophisticated
long-range-corrected functionals: 
van Leeuwen-Baerends (LB94) \cite{AgAu-alloy:2011}, 
Gritsenko-van Leeuwen-van Lenthe-Baerends (GLLB) \cite{gpaw-prop:2015} and 
long-range corrected PBE (LC-$\omega$PBE)
\cite{lrc-funct:2009,lrc-funct:2010,small-ag-cu:2014}.

Quantitatively assessing the position of the low-frequency resonance, we
can compare our results with several theoretical calculations.
For Ag$_{13}$ cluster, we can find the values (in eV)
3.5 (PBE) \cite{ag13-in-ar:2008},
3.5 (LC-$\omega$PBE) \cite{small-ag-cu:2014},
3.2 (LDA) \cite{Ag13-lda-analysis:2008}, 
3.7 (LDA) \cite{xc-contr-ag13-ag55:2010} and 
3.2 (PBE) \cite{Rao201350:2013}, while we get the maximum at 3.63 eV.
For Ag$_{55}$ icosahedral cluster, we can find values 
3.5 (LDA) \cite{xc-contr-ag13-ag55:2010} and 
4.2 (GLLB) \cite{gpaw-prop:2015} eV, while we get the maximum at 3.71 eV.
For Ag$_{147}$ icosahedral cluster, we can find values
3.8 (GLLB) \cite{gpaw-prop:2015},
3.2 (PBE) \cite{octopus-ag-au:2013},
4.5 (LB94) \cite{AgAu-alloy:2011}, 
while we get the maximum at 3.5 eV.
For Ag$_{309}$ icosahedral cluster, 
we can find values 3.7 (GLLB) \cite{gpaw-prop:2015},
3.50 (polarizability interaction model) \cite{pim:2010},
while we get the maximum at 3.37 eV.
Finally, for Ag$_{561}$ icosahedral cluster, there is a value 3.65 eV
from \cite{gpaw-prop:2015}, while we get 3.25 eV.
Summarizing, we see that the deviations between our calculations and 
other calculations that utilize the LDA or GGA functionals do not exceed
0.2 eV and might be due to differences in the used basis sets or pseudo-potentials.
The larger deviations of about 0.4 eV, we have with LRC functionals.

Hollow structures have been produced and characterized \cite{exp-shells:2002,
exp-gold-shells:2003,
exp-shells:2005,rev-plasmonics:2010,exp-shells:2013,sigmaaldrich:2015}.
Experimental evidence supports our finding that the hollow structures have
lower frequencies of the main plasmonic resonance.
However, the measurements were performed
with relatively large and thick shells (at least one order of 
magnitude larger than those considered here) and a quantitative comparison of 
our results with experiments is hardly possible.

From the theoretical side, there are several estimations for hollow metallic
particles available. Firstly, the solution of Maxwell equations in the 
quasi-static limit 
for a hollow spherical particle is known \cite{Lucas:1994} and the effect 
of removing interior part of the sphere is a red shift of the plasmon frequency.
As in the case of compact spheres, the resonance frequencies do not depend 
on the size of the system, 
but only on the ratio of the inner and outer radius of the 
sphere. Secondly, there are atomistic TDDFT calculations for icosahedral shells
of up to size of 6 layers available \cite{Weissker:2014, Stener-shells:2014}.
These calculations show red shifted plasmonic resonances for hollow clusters
compared to the filled ones. Quantitatively comparing the position of the 
low-frequency resonance, we get a fair agreement. For instance, for 
Ag$_{92}$ (S4L1 shell) we can find 
3.85 (LB94) \cite{Stener-shells:2014} and 
2.8 (PBE) \cite{Weissker:2014} eV, 
while we get 2.97 eV. 
For Ag$_{12}$, Ag$_{42}$, Ag$_{162}$ and
Ag$_{252}$ (S2L1, S3L1, S5L1 and S6L1 shells) we extract from 
\cite{Stener-shells:2014} 4.0, 4.35, 3.5 and 3.2 eV,
while we get 4.17, 3.13, 2.8 and 2.63 eV, respectively. The discrepancies
are sizeable, but can be explained by the differences of the functional
(LB94 versus WC) and the geometry relaxations (ideal symmetric with the 
nearest-neighbor distance fixed at a equilibrium bulk value 2.89 \AA{}
versus fully relaxed geometries with no symmetry imposed in our case).
The effect of geometry relaxations is estimated in subsection 
\ref{ss:shells-results}: it has a minor importance compared to
the influence of DFT functional and does blueshift the resonances of 
the ideal structures with respect to those of relaxed ones.

The response of other silver made structures with effectively reduced
dimensionality: rods \cite{sonnichsen:2001,prl-au-nanorods-red-dumping:2002,
octopus-ag-au:2013,small-ag-cu:2014} and
plateletts \cite{exp-shells:2005,prolate-oblate:2013} also 
exhibit red shifted resonance frequencies of the plasmon excitations
with respect to spherical or quasi-spherical clusters of similar sizes.

\section{Conclusion\label{s:conclusion}}

We studied the optical response of silver clusters of icosahedral symmetry 
and their hollow counterparts by means of quantum mechanical, atomistic
methods (linear response TDDFT within LCAO). The applied iterative methods
allowed for comparatively fast calculations (about 1 day of walltime) of 
compact clusters containing up to 561 atoms and hollow clusters
of up to 868 atoms. We have found that the plasmonic resonance of
silver clusters depends on the size and morphology of clusters. 
Namely, the frequency of maximal absorption of the
icosahedral clusters that contain three and more atom layers is inversely
proportional to the cluster diameter. Moreover, the single-layered
shells show a sizeable red shift of the resonance frequencies, which 
quickly becomes negligible as the thickness of the shells increases.
Both observations are compatible with experimental findings, previous
calculations, and can be partially understood within classical electrodynamics.
Furthermore, both observations are valid also for the high-frequency,
interband plasmon which was not widely studied so far.

From a methodological point of view, we presented recent developments
of our iterative technique including a realization of an atom-centered
product basis and various analysis tools. The iterative method used here
\cite{iter_method}, particularly with the speedup allowed by the use of an
atom-centered auxiliary basis to express the orbital products,
is advantageous in many respects that are discussed in the paper. 
In particular, the frequency-range selectivity is useful in calculations
of the Raman response, for which one needs the response 
in a very narrow spectral range.
Moreover, the current implementation of the response times 
vector operation can be easily optimized to be less memory demanding.
This optimization will allow to extend the number of treated atoms by an
order of magnitude, and even further if full MPI parallelization 
is implemented.

\section*{Acknowledgments}

This work is supported, in part, by the ORGAVOLT (\textsc{ORGAnic}
solar cell \textsc{VOLTage} by numerical computation) Grant ANR-12-MONU-0014-02
of the French {Agence Nationale de la Recherche (ANR) 2012 Programme Mod\`eles Num\'eriques}.
Federico Marchesin, Peter Koval and Daniel S\'anchez-Portal 
acknowledge support from the {Deutsche Forschungsgemeinschaft (DFG)}
through the \textsc{SFB}1083 project, the Spanish \textsc{MINECO}
MAT2013-46593-C6-2-P project, the Euroregion Aquitaine-Euskadi
program and from the Basque {Departamento de Educaci\'on}, \textsc{UPV/EHU}
(Grant No.\ IT-756-13). Peter Koval acknowledges financial support
from the Fellows Gipuzkoa program of the {Gipuzkoako Foru Aldundia} through
the FEDER funding scheme of the European Union.

\appendix

\section{Dominant basis set symmetries 
contributing to the optical polarizability\label{a:analysis}}

The prior analysis in section \ref{ss:analysis} has been formulated for a 
simple, physically motivated splitting of the interacting polarizability 
in terms of angular momentum of the occupied states and in terms 
of the atomic contributions to the optical polarizability.
In this section, we focus on a more technical analysis of 
the interacting polarizability in terms of the products of 
atomic-orbital functions, and in terms of the product basis functions
contributing to the resulting induced screened density change in the cluster.

Similarly to the analysis tools considered in section \ref{ss:analysis}, the 
total interacting polarizability can be split into sums with fixed 
angular momentum of the product functions $F^{\mu}(\bm{r})$.
This is so because the product functions are constructed as a linear
combination of the products of atomic orbitals separately for
each possible angular momentum of the product \cite{df-pk:2009}
and, therefore, the product function $\mu$ carries a well-defined
angular momentum $l_{\mu}$. We can write the polarizability 
as a sum over angular momenta of the product basis 
$\alpha(\omega) = \sum_{l} \alpha_l(\omega)$ with 
a product angular momentum resolved polarizability 
$\alpha_l(\omega)$ given by

\begin{equation}
\alpha_l(\omega) = d^{\mu} \delta_{l_{\mu}, l} \delta n_{\mu}(\omega),
\label{pol_prd_ang_mom}
\end{equation}
where the dipole moments $d^{\mu}$ refer to a global origin
of the coordinate system. The dipole moments are defined for product
functions $F^{\mu}(\bm{r})$ that are centered on atoms
\begin{equation}
\bm{d}^{\mu} = \int F^{\mu}(\bm{r}-\bm{R}_{\mu}) \bm{r} d^3r =
\int F^{\mu}(\bm{r}) \bm{r} d^3r +\bm{R}_{\mu} \int F^{\mu}(\bm{r}) d^3r. 
\label{dip-mom}
\end{equation}
The last equation makes it apparent that only angular momentum 
$l_{\mu}=0$ and $l_{\mu}=1$ can contribute to the dipole polarizability.

Yet another type of analysis involves the angular momentum 
symmetry of the atomic orbitals, the products of which 
expand the density change $\delta n(\omega,\bm{r})$.
Namely, the product function $F^{\mu}(\bm{r})$ is expressed 
in terms of a linear combination of the products of atomic orbitals
(\ref{df-constr}). Therefore, we can immediately write the 
interacting polarizability $\alpha(\omega)$ as a sum over
atomic orbital angular momentum sub-sums
$\alpha(\omega) = \sum_{l_1, l_2} \alpha_{l_1, l_2}(\omega)$
with the partial polarizability $\alpha_{l_1, l_2}(\omega)$  given by
\begin{equation}
\alpha_{l_1, l_2}(\omega) = 
\delta_{l_a,l_1} d^{ab} \delta_{l_b,l_2} \Lambda_{ab}^{\mu} \delta n_{\mu}(\omega).
\label{pol_orb_ang_mom}
\end{equation}
Here the dipole matrix elements between atomic orbitals is used 
$d^{ab} = \int f^{a}(\bm{r}-\bm{R}_a)\bm{r}f^{b}(\bm{r}-\bm{R}_b)dr$. 
Thus, recalling that the atom-centered functions are constructed
from local on-site products, we see that the total polarizability
can be expressed in terms of the on-site atomic orbitals. 
From the one side, this reveals a certain arbitrarity of 
the proposed separation, but from the other side, 
it might be useful for developing simplified models.
Because the global origin and the atom centers do not generally
coincide, there are no dipole selection rules in the matrix-elements
$d^{ab}$. Namely, analogously to the case, of dipole moments
(\ref{dip-mom}), the matrix elements $d^{ab}$  will depend on the 
overlap between orbitals

\begin{equation}
\bm{d}^{ab} = \int f^{a}(\bm{r})\bm{r}f^{b}(\bm{r})dr +
\bm{R} \int f^{a}(\bm{r}) f^{b}(\bm{r})dr,
\end{equation}
where $\bm{R} = \bm{R}_a = \bm{R}_b$. The matrix element $d^{ab}$
is zero if both term in the last equation are zero. This is 
always the case for $sd$ channels ($l_a=0, l_b=2$), but not
generally for $ss$ and $pp$ symmetries.

\begin{figure}[htbp]
\includegraphics[width=14cm]{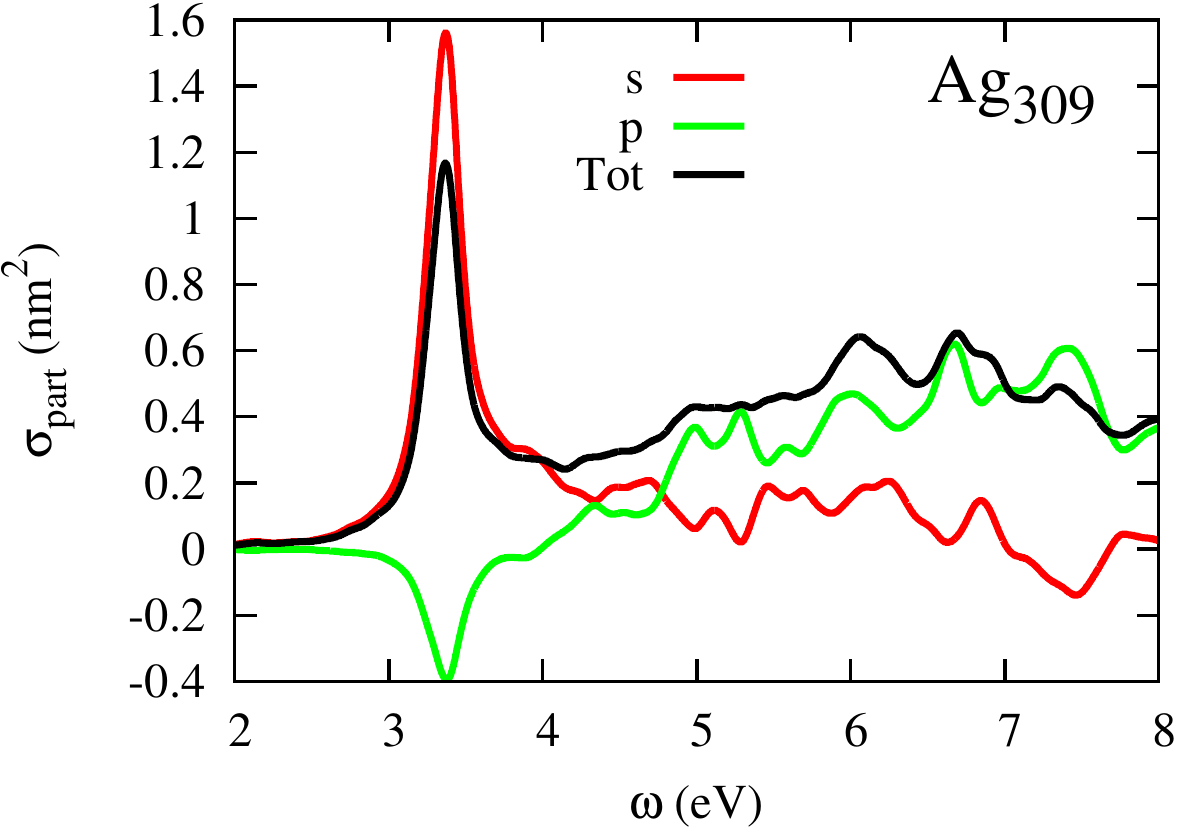}
\caption{\label{f:prd_ang_mom}Analysis of the 
product-function angular-momentum contributions to the absorption
cross section of the Ag$_{309}$ cluster.
The total absorption cross-section is also plotted for comparison.}
\end{figure}

In figure \ref{f:prd_ang_mom}, we present the 
analysis of the product-function angular momentum 
contributions to the optical absorption cross section.
The partial cross sections corresponding to the 
partial polarizability (\ref{pol_prd_ang_mom}) are plotted
together with the total absorption cross section for 
Ag$_{309}$ cluster. The contribution of the angular 
momentum higher than $p$ is strictly zero. The contribution 
of $s$-symmetric product functions dominates in the 
cross section, while the contribution of $p$-symmetric
functions is negative for the low-frequency resonance
in the range 3--4 eV and positive in the frequency 
range 5--8 eV.

\begin{figure}[htbp]
\includegraphics[width=14cm]{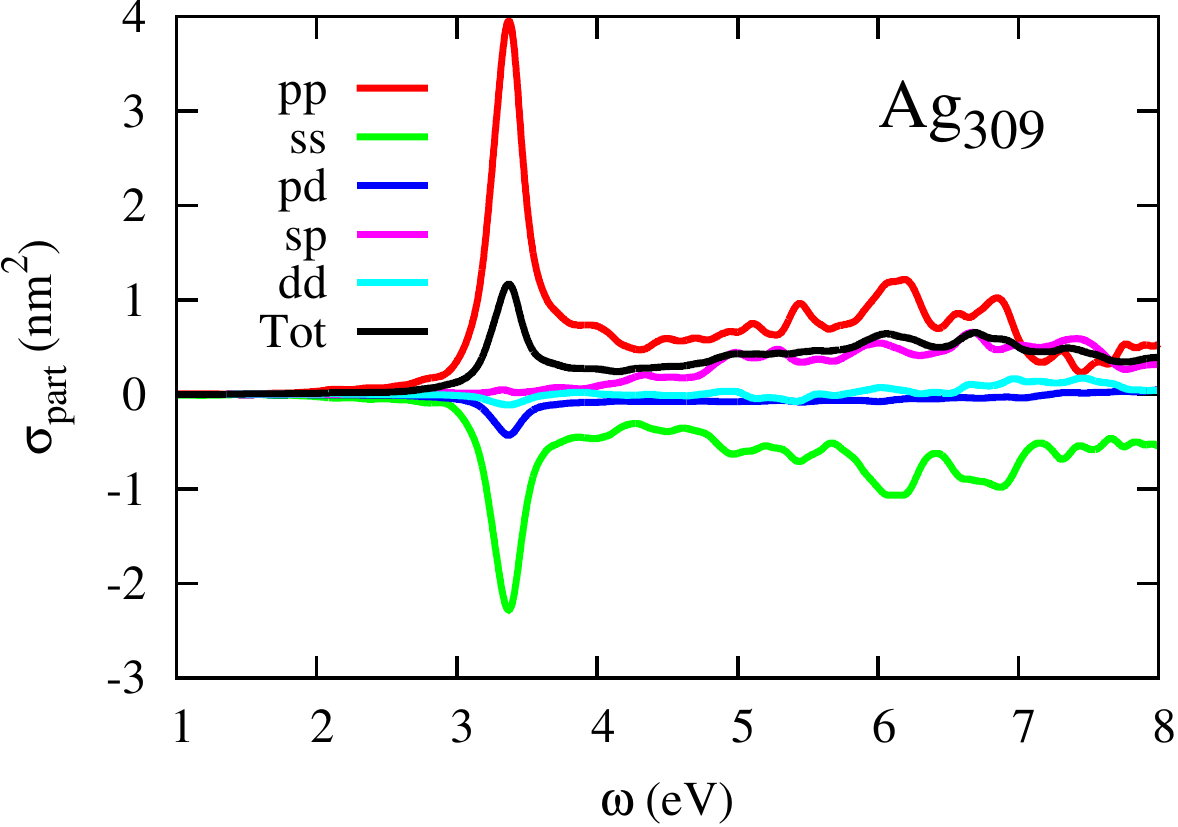}
\caption{\label{f:orb_ang_mom}Analysis of atomic orbital angular
momenta of the products of atomic-orbitals contributions 
to the optical absorption cross-section
of the Ag$_{309}$ cluster. The total absorption cross-section is
also plotted for comparison.}
\end{figure}

In the figure \ref{f:orb_ang_mom}, we plot the partial cross
sections computed from the orbital angular-momentum-resolved
polarizability (\ref{pol_orb_ang_mom}) for the 5-layered
icosahedral cluster Ag$_{309}$. The sum of the partial cross sections
is also shown for comparison. We can clearly
appreciate the contribution of $p$-orbitals to the total absorption
cross section. Namely, the contribution of the $pp$ angular-momentum 
channel is the most significant.  
The contributions have different signs and, for example,
$pp$-products give rise to a partial cross section 
that is about 3 times larger than the total cross section.
The second most important type of products is of $ss$ type.
The partial cross section of $ss$ type is negative in the whole 
frequency range we computed. The absolute value of $ss$ partial
cross section is about 1.7 times larger than the total cross section.
The other non-zero angular-momentum channels do not contribute 
significantly to the total absorption cross section. Therefore,
a linear combination of $pp$ and $ss$ products of atomic orbitals
seems to be able to provide a reasonably accurate 
description of the total absorption cross section 
in the frequency range 0--10~eV and even better so
in the frequency range 0--4~eV.

\section{GGA interaction kernel\label{a:gga-kernel}}

In the generalized gradient approximation (GGA) of DFT functional, the 
xc energy density $\epsilon$ explicitly depends 
on the charge density $n(\bm{r})$ and its gradient $\nabla n(\bm{r})$
\begin{equation}
E = \int \epsilon( n(\bm{r}), \nabla n(\bm{r}) ) d^3 r.
\label{energy}
\end{equation}
In the Kohn-Sham formulation of DFT we need the functional derivative 
of the energy with respect to the electron density
\begin{equation}
v(\bm{r}) \equiv \frac{\delta E }{\delta n(\bm{r}) }=
\frac{\partial \epsilon}{\partial n} - 
\nabla \frac{\partial \epsilon}{\partial \nabla n}.
\label{pot1}
\end{equation}
and in the linear response TDDFT we need also a 
second variational derivative of the energy 
with respect to the electronic density 
\begin{equation}
f(\bm{r},\bm{r}') \equiv \frac{\delta v(\bm{r}) }{\delta n(\bm{r}') }.
\label{kernel}
\end{equation}
While the calculation of the KS potential in GGA is widely discussed
in the literature \cite{white-bird:1994,bylander:1997,soler-gga:2001},
the TDDFT kernel (\ref{kernel}) is less commonly discussed 
\cite{Gaiduk:1996,Amos:2000} and we find it convenient to 
state explicitly the equations necessary for the implementation 
of our iterative TDDFT method.

\subsection{Variational derivative of potential}
 
In linear response TDDFT one needs a first variational derivative of the 
xc potential with respect to density (\ref{kernel}).
The potential (\ref{pot1}) is in fact a functional of the density 
and its gradient. Computing the variation of the potential one gets 

\begin{eqnarray}
\delta v(\bm{r}) = 
\frac{\partial^2 \epsilon }{\partial n \partial n} \delta n +
\frac{\partial^2 \epsilon }{\partial n \partial \nabla_i n} \nabla_i \delta n
&-& 
\nabla_i\left( \frac{\partial^2 \epsilon }{\partial n \partial \nabla_i n} 
\delta n \right) \nonumber \\
&-&
\nabla_i\left( \frac{\partial^2 \epsilon }{\partial \nabla_k n \partial \nabla_i n} 
\nabla_k \delta n \right). 
\label{var_pot1}
\end{eqnarray}
In order  to ``exchange'' the gradients of density variation $\nabla_i \delta n$
and $\nabla_k\nabla_i \delta n$ for the density variation $\delta n$,
we represent the last equation in form of an integral with a
$\delta$-function
$\delta v(\bm{r}') = \int \delta(\bm{r}'-\bm{r})\delta v(\bm{r})  d^3 r$
and apply the integration by parts whenever necessary.
Finally we can get \cite{Nazarov-priv-commun:2015}
\begin{eqnarray}
\delta v(\bm{r}') &=& \int 
\delta(\bm{r}'-\bm{r}) \delta n \frac{\partial^2 \epsilon }{\partial n \partial n} 
-
\delta(\bm{r}'-\bm{r}) \delta n 
\nabla_i \left[\frac{\partial^2 \epsilon }{\partial n \partial \nabla_i n}\right] 
\nonumber\\
&-& \delta n \nabla_k
\left[\nabla_i \left[\delta(\bm{r}'-\bm{r})\right]
\frac{\partial^2 \epsilon }{\partial \nabla_k n \partial \nabla_i n} \right]
 d^3 r. 
\label{var_pot5}
\end{eqnarray}

By virtue of the kernel definition (\ref{kernel}), the kernel GGA reads
\cite{Nazarov-prl:2011}
\begin{eqnarray}
\label{kernel-1}
f(\bm{r},\bm{r}') = 
\delta(\bm{r}-\bm{r}')\frac{\partial^2 \epsilon }{\partial n \partial n} 
&-&
\delta(\bm{r}-\bm{r}') 
\nabla_i \left[\frac{\partial^2 \epsilon }{\partial n \partial \nabla_i n}\right] 
\nonumber\\
&-&
\nabla_k
\left[\nabla_i \left[\delta(\bm{r}-\bm{r}')\right]
\frac{\partial^2 \epsilon }{\partial \nabla_k n \partial \nabla_i n} \right].
\end{eqnarray}

\subsection{Computation of GGA matrix elements of potential}

In DFT, we need matrix elements of the xc potential (\ref{pot1})
\begin{equation}
v^{ab} = 
\int f^a(\bm{r}) \frac{\partial \epsilon}{\partial n} f^b(\bm{r}) d^3r - 
\int f^a(\bm{r}) \nabla \frac{\partial \epsilon}{\partial \nabla n} f^b(\bm{r}) d^3r
\label{pot1_me}
\end{equation}
between the atomic orbitals $f^a(\bm{r})$. 
Because we use \textsc{LIBXC} library \cite{libxc:2012}, we need 
to work with derivatives of the xc energy density
with respect to the square of the density gradient, 
i.e. in terms of $\sigma \equiv (\nabla n)^2$.
After applying the chain rule, we transform the last formula to 
\begin{equation}
v^{ab} = 
\int f^a(\bm{r}) \frac{\partial \epsilon}{\partial n} f^b(\bm{r}) d^3r - 
2\int f^a(\bm{r}) \nabla_i [\frac{\partial \epsilon}{\partial \sigma} \nabla_i n]
f^b(\bm{r}) d^3r,
\label{pot2_me}
\end{equation}
Here and in the following the summation over repeated Cartesian indices 
($i$ and $k$) is understood.
The expression (\ref{pot2_me}), 
similar to that in equation (\ref{pot1_me}), is inconvenient
in a numerical calculation because one must 
compute derivatives of the quantities
delivered by \textsc{LIBXC} library ($\frac{\partial \epsilon}{\partial \nabla n}$
or $\frac{\partial \epsilon}{\partial \sigma}\nabla n$ ). 
However, with one more 
integration by parts, we can transform the expression (\ref{pot2_me})
into a more suitable form
\begin{eqnarray}
v^{ab} = &
\int f^a(\bm{r})\  &\frac{\partial \epsilon}{\partial n} 
\ 
f^b(\bm{r}) d^3r \nonumber\\  
&+ 2 \int \nabla_i [f^a(\bm{r})] \nabla_i n \ 
&\frac{\partial \epsilon}{\partial \sigma}\ 
f^b(\bm{r})\, d^3r \nonumber\\ 
&+  2 \int f^a(\bm{r}) \ 
&\frac{\partial \epsilon}{\partial \sigma}\ 
\nabla_i [f^b(\bm{r})] \nabla_i n\, d^3r.
\label{pot3_me}
\end{eqnarray}
This form is advantageous because it is easier to calculate the
gradients of the basis functions, the form of which is known and does
not change with the type of the xc functional.
Moreover, the last expression uses the same ingredients that
are necessary to compute the matrix elements of GGA kernel.

\subsection{Computation of GGA matrix elements of kernel}

In our implementation of the linear response TDDFT, 
we need the matrix elements of the xc kernel (\ref{kernel-1}) defined by 
$K^{\mu\nu} = \int F^{\mu}(\bm{r}) f(\bm{r},\bm{r}') F^{\nu}(\bm{r}') d^3 r d^3r'$.
Exercising the same approach as for the potential in the previous 
subsection, we obtain for matrix elements of GGA kernel

\begin{eqnarray}
K^{\mu\nu} = 
\int F^{\mu}(\bm{r})\ 
&\frac{\partial^2 \epsilon }{\partial n^2}&\ 
F^{\nu}(\bm{r}) d^3r 
\nonumber \\
+ 2\int \nabla_i[ F^{\mu}(\bm{r}) ] \nabla_i[n]\ 
&\frac{\partial^2 \epsilon }{\partial n \partial \sigma}&\ 
F^{\nu}(\bm{r}) d^3r 
\nonumber \\ 
+ 2\int F^{\mu}(\bm{r})\ 
&\frac{\partial^2 \epsilon }{\partial n \partial \sigma}&\ 
\nabla_i[F^{\nu}(\bm{r})]\nabla_i[n] d^3r \nonumber\\
+ 4\int \nabla_k[F^{\mu}(\bm{r})] \nabla_k[n]\ 
& \frac{\partial^2 \epsilon }{\partial \sigma^2}&\ 
\nabla_i [F^{\nu}(\bm{r})]\nabla_i[n] d^3r \nonumber\\ 
+2 \int \nabla_k[F^{\mu}(\bm{r})]\ 
&\frac{\partial \epsilon }{\partial \sigma}&\  \delta_{ik}
\nabla_i[F^{\nu}(\bm{r})] d^3r.
\label{kernel_me}
\end{eqnarray}
These matrix elements are computed with standard 
integration methods in quantum chemistry 
\cite{Delley:1990, Krack:1998}.

\clearpage

\bigskip
\bigskip
\bigskip
\bibliographystyle{iopart-num}
\bibliography{ms}

\end{document}